%% file: main.tex
\documentclass[11pt, oneside]{article}

\input{preamble.tex}

\usepackage{credits}
\usepackage{orcidlink}
\usepackage{hyperref}
\usepackage{multirow}
\usepackage{threeparttable}

\usepackage{xcolor}
\usepackage{booktabs}
\usepackage{xltabular}
\usepackage{array}
\usepackage{ragged2e}
\usepackage{tcolorbox}
\usepackage{svg}

\usepackage{listings}
\usepackage{tcolorbox}
\tcbuselibrary{listings,breakable,xparse,skins}

\definecolor{bg}{gray}{0.95}
\DeclareTCBListing{mintedbox}{O{}m!O{}}{%
  breakable,
  listing engine=listings,
  listing only,
  listing options={
    language=Python, 
    basicstyle=\small\ttfamily,
    keywordstyle=\color{blue},
    stringstyle=\color{red},
    commentstyle=\color{gray},
    numbers=left,
    numberstyle=\tiny,
    breaklines=true,
    #1
  },
  boxsep=0pt, left=25pt, right=0pt, top=3pt, bottom=3pt,
  arc=5pt, leftrule=0pt, rightrule=0pt, bottomrule=2pt, toprule=2pt,
  colback=bg, colframe=red!75!black,
  enhanced,
  overlay={\begin{tcbclipinterior}\fill[red!50!white] (frame.south west) rectangle ([xshift=20pt]frame.north west);\end{tcbclipinterior}},
  #3}

\title{\textsf{An autonomous living database for perovskite photovoltaics}}

\input{authors}

\begin{document}

\maketitle

\begin{abstract}
    \noindent Scientific discovery is severely bottlenecked by the inability of manual curation to keep pace with exponential publication rates. This creates a widening knowledge gap. This is especially stark in photovoltaics, where the leading database for perovskite solar cells has been stagnant since 2021 despite massive ongoing research output. Here, we resolve this challenge by establishing an autonomous, self-updating living database (\perla). Our pipeline integrates large language models with physics-aware validation to extract complex device data from the continuous literature stream, achieving human-level precision ($>90\%$) and eliminating annotator variance. By employing this system on the previously inaccessible post-2021 literature, we uncover critical evolutionary trends hidden by data lag: the field has decisively shifted toward inverted architectures employing self-assembled monolayers and formamidinium-rich compositions, driving a clear trajectory of sustained voltage loss reduction. \perla transforms static publications into dynamic knowledge resources that enable data-driven discovery to operate at the speed of publication.
\end{abstract}

\input{sections/introduction}
\input{sections/results}

\input{sections/conclusions}
\input{sections/methods}

\section*{Acknowledgements}
The work of K.M.J.\ has been supported by the Carl-Zeiss Stiftung, Intel and Merck via the AWASES Project, and a Google Research Scholar Award.
Part of the work has been supported by the Helmholtz Association within the framework of the Helmholtz Foundation Model Initiative (project SOL-AI).
This work was also supported by the NFDI consortium FAIRmat - Deutsche Forschungsgemeinschaft (DFG) - Project 460197019 and by the SolMates project, funded by the European Union’s Horizon Europe research and innovation program under grant agreement No 101122288. We thank Sreekanth Kunchapu for code contributions and Philippe Holzhey for useful discussions. We also thank Berfin Güner for assistance with the design of Figure 1. 

\section*{Author contributions}

\footnotesize
\insertcredits
\normalsize 

\section*{Conflicts of interests}
The authors declare no conflicts of interest.

\section*{Data and code availability}
Code is available on GitHub (\url{https://github.com/FAIRmat-NFDI/perla}) with documentation hosted on \url{https://fairmat-nfdi.github.io/perla}.

The newly extracted data alongside the legacy perovskite database data can be explored in NOMAD at \url{https://nomad-lab.eu/prod/v1/gui/search/perovskite-solar-cells-database?upload_name=PERLA%20Seed&upload_name=PERLA%20Bot}.

\section*{Declaration of Generative AI and AI-assisted technologies in the writing process}
During the preparation of this work, the authors used Anthropic’s Claude and Google's Gemini models to improve language and readability. After using this service, the authors reviewed and edited the content as needed and take full responsibility for the content of the publication.

\printbibliography 

\clearpage
\appendix
\input{sections/appendix}

\end{document}

%% file: preamble.tex
\usepackage[english]{babel}
\usepackage{geometry}  
\usepackage{graphicx}      
\usepackage{tabularx}      
\usepackage{amssymb}
\usepackage{amsmath}
\usepackage{pdflscape}
\usepackage{booktabs}
\usepackage{xltabular}
\usepackage[hyphens]{url}
\usepackage[hidelinks]{hyperref}
\hypersetup{breaklinks=true}

\usepackage{mdframed}
\usepackage[backend=biber,style=chem-acs, doi=false,isbn=false, date=year, url=false, autopunct=true]{biblatex}
\makeatletter
\renewbibmacro*{textcite}{%
  \iffieldequals{namehash}{\cbx@lasthash}
    {\usebibmacro{cite:comp}}
    {\usebibmacro{cite:dump}%
     \ifbool{cbx:parens}
       {\bibclosebracket\global\boolfalse{cbx:parens}}
       {}%
     \iffirstcitekey
       {}
       {\textcitedelim}%
     \usebibmacro{cite:init}%
     \ifnameundef{labelname}
       {\printfield[citetitle]{labeltitle}}
       {\printnames{labelname}}%
     \addspace
     \ifnumequal{\value{citecount}}{1}
       {\usebibmacro{prenote}}
       {}%
     \mkbibsuperscript{\usebibmacro{cite:comp}}%
     \stepcounter{textcitecount}%
     \savefield{namehash}{\cbx@lasthash}}}
\makeatother

\usepackage{csquotes}
\usepackage{cleveref}
\usepackage{siunitx}
\usepackage{geometry}
\usepackage{multirow}
\usepackage{mhchem}
\usepackage[dvipsnames,table,xcdraw]{xcolor}
\definecolor{tumblue}{rgb}
{0.6431372549,0.2196078431,0.1882352941}

\usepackage{xspace}
\usepackage{caption}
\usepackage{subcaption}
\usepackage{authblk} 

\usepackage{marvosym}

\usepackage{sectsty} 

\sectionfont{\color{tumblue}\selectfont\sffamily}
\subsectionfont{\color{tumblue}\selectfont\sffamily}
\subsubsectionfont{\color{tumblue}\selectfont\sffamily}
\paragraphfont{\selectfont\sffamily}

\IfFileExists{./fonts/Heliotrope-Bold.otf}{
    \usepackage{fontspec}
    \setmainfont{Heliotrope}[Extension = .otf, UprightFont = *-Regular, ItalicFont = *-Italic, BoldFont=*-Bold, Path = ./fonts/]
}{
    \usepackage{biolinum}
    
}

\usepackage{lipsum}
\usepackage[labelfont=bf]{caption}
\usepackage[nonumberlist,acronym]{glossaries}
\setacronymstyle{long-short}
\input{acronyms}

\makenoidxglossaries

 \usepackage{microtype}
 \usepackage{fontawesome}

\definecolor{linkcolor}{RGB}{0, 0, 0}

\newcommand{\perla}{PERLA\xspace}

\usepackage{tikz}
\usetikzlibrary{shapes.geometric, arrows.meta, positioning, fit, backgrounds}

%% file: acronyms.tex
\newglossaryentry{dft}{
    type=\acronymtype, 
    name={DFT}, 
    description={density functional theory},
}

\newglossaryentry{mae}{
    type=\acronymtype, 
    name={MAE}, 
    description={mean absolute error},
}

\newglossaryentry{rmse}{
    type=\acronymtype, 
    name={RMSE}, 
    description={root mean squared error},
}

\newglossaryentry{llm}{
    type=\acronymtype, 
    name={LLM}, 
    description={large language model},
}

\newglossaryentry{ml}{
    type=\acronymtype, 
    name={ML}, 
    description={machine learning},
}

\newglossaryentry{mof}{
    type=\acronymtype, 
    name={MOF}, 
    description={metal-organic framework},
}

\newglossaryentry{casp}{
    type=\acronymtype, 
    name={CASP}, 
    description={Critical Assessment of Structure Prediction},
}

\newglossaryentry{harking}{
    type=\acronymtype, 
    name={HARKing}, 
    description={Hypothesizing After the Results are Known},
}

\newglossaryentry{api}{
    type=\acronymtype, 
    name={API}, 
    description={application programming interface},
}

\newglossaryentry{tco}{
    type=\acronymtype, 
    name={TCO}, 
    description={transparent conductive oxide},
}

\newglossaryentry{mlff}{
    type=\acronymtype, 
    name={MLFF}, 
    description={machine learning force field},
}

\newglossaryentry{md}{
    type=\acronymtype, 
    name={MD}, 
    description={molecular dynamics},
}

\newglossaryentry{roc}{
    type=\acronymtype, 
    name={ROC}, 
    description={receiver operating characteristic},
}

\newglossaryentry{pr}{
    type=\acronymtype, 
    name={PR}, 
    description={precision recall},
}

%% file: authors.tex
\author[1]{Sherjeel~Shabih~\orcidlink{0009-0007-4392-5918}}
\author[1]{Hampus~Näsström~\orcidlink{0000-0002-3264-1692}}
\author[1]{Sharat~Patil}
\author[2]{Asmin~Askin~\orcidlink{0000-0002-3381-7258}}
\author[2]{Keely~Dodd-Clements~\orcidlink{0009-0006-7923-0600}}
\author[8]{Jessica~Helisa~Hautrive~Rossato~\orcidlink{0000-0003-1644-5680}}
\author[8]{Hugo~Gajardoni~de~Lemos~\orcidlink{0000-0003-4525-9967}}
\author[2]{Yuxin~Liu~\orcidlink{0009-0009-3966-9901}}
\author[2]{Florian~Mathies~\orcidlink{0000-0002-8950-3901}}
\author[2]{Natalia~Maticiuc~\orcidlink{0000-0002-5654-3560}}
\author[2]{Rico~Meitzner}
\author[2]{Edgar~Nandayapa~\orcidlink{0000-0002-5248-7143}}
\author[2]{Juan~José~Patiño~López~\orcidlink{0000-0002-8950-3901}}
\author[1]{Yaru~Wang}
\author[1]{Lauri~Himanen~\orcidlink{0000-0002-3130-8193}}
\author[1,2]{Eva~Unger~\orcidlink{0000-0002-3343-867X}}
\author[3]{T.~Jesper~Jacobsson~\orcidlink{0000-0002-4317-2879}}
\author[1, \Letter, $\star$]{José~A.~Márquez~\orcidlink{0000-0002-8173-2566}}
\author[4,5,6,7, \Letter, $\star$]{Kevin~Maik~Jablonka~\orcidlink{0000-0003-4894-4660}}

\affil[1]{Department of Physics and CSMB, Humboldt-Universität zu Berlin, Berlin, Germany}

\affil[2]{HySPRINT Innovation Lab, Helmholtz-Zentrum Berlin für Materialien und Energie GmbH, Berlin, Germany}

\affil[3]{Department of Physics, Chemistry and Biology (IFM), Linköping University, Linköping, Sweden}

\affil[4]{Laboratory of Organic and Macromolecular Chemistry (IOMC), Friedrich Schiller University Jena, Humboldtstrasse 10, 07743 Jena, Germany}

\affil[5]{Center for Energy and Environmental Chemistry Jena (CEEC Jena), Friedrich Schiller University Jena, Philosophenweg 7a, 07743 Jena, Germany}

\affil[6]{Helmholtz Institute for Polymers in Energy Applications Jena (HIPOLE Jena), Lessingstrasse 12--14, 07743 Jena, Germany}

\affil[7]{Jena Center for Soft Matter (JCSM), Friedrich Schiller University Jena, Philosophenweg 7, 07743 Jena, Germany}

\affil[8]{UNESP -- Universidade Estadual Paulista ``Júlio de Mesquita Filho'', Bauru, São Paulo, Brazil}

\affil[$\star$]{These authors contributed equally (ordering was determined using \texttt{import random; a = ['Pepe', 'Kevin']; random.choice(a)}).}

\affil[\Letter]{\texttt{jose.marquez@physik.hu-berlin.de} and \texttt{mail@kjablonka.com}}

\credit{Sherjeel~Shabih}{0,1,1,0,1,1,0,0,1,0,1,1,0,1}

\credit{Hampus~Näsström}{0,1,0,0,0,0,0,0,1,0,0,0,0,1}
\credit{Sharat~Patil}{0,1,0,0,0,0,0,0,1,0,0,1,0,1}
\credit{Asmin~Askin}{0,1,0,0,0,0,0,0,0,0,0,0,0,1}
\credit{Keely~Dodd-Clements}{0,1,0,0,0,0,0,0,0,0,0,0,0,1}
\credit{Jessica~Helisa~Hautrive~Rossato}{0,1,0,0,0,0,0,0,0,0,0,0,0,1}
\credit{Hugo~Lemos}{0,1,0,0,0,0,0,0,0,0,0,0,0,1}
\credit{Yuxin~Liu}{0,1,0,0,0,0,0,0,0,0,0,0,0,1}
\credit{Florian~Mathies}{0,1,0,0,0,0,0,0,0,0,0,0,0,1}
\credit{Natalia~Maticiuc}{0,1,0,0,0,0,0,0,0,0,0,0,0,1}
\credit{Rico~Meitzner}{0,1,0,0,0,0,0,0,0,0,0,0,0,1}
\credit{Edgar~Nandayapa}{0,1,0,0,0,0,0,0,0,0,0,0,0,1}
\credit{Juan~José~Patiño~López}{0,1,0,0,0,0,0,0,0,0,0,0,0,1}
\credit{Yaru~Wang}{0,0,0,0,0,0,0,0,1,0,0,0,0,1}
\credit{Lauri~Himanen}{0,0,0,0,0,0,0,0,1,0,0,0,0,1}
\credit{Eva Unger}{0,0,0,0,0,0,0,1,0,0,0,0,0,1}
\credit{Jesper~T.~Jacobsson}{0,1,0,0,0,0,0,0,0,0,0,0,0,1}
\credit{José~A.~Márquez}{1,1,1,1,1,1,1,1,1,1,1,1,1,1}
\credit{Kevin Maik Jablonka}{1,1,1,1,1,1,1,1,1,1,1,1,1,1}

%% file: sections/introduction.tex
\section{Introduction}
Perovskite solar cell research now produces results faster than scientists can systematically learn from them. 
Hundreds of papers appear each month, reporting device efficiencies and processing protocols for different architectures.\autocite{extraction_review} 
While this knowledge exists and would be essential for systematic advancement of the field, it is mostly trapped in papers.

The Perovskite Database (PDB) demonstrated what structured curation makes possible.\autocite{perovskite_db, Unger_Jacobsson_2022} 
By manually extracting approximately 100 parameters from 42,400 devices across 15,000 papers, the project established FAIR data standards\autocite{fair} and enabled population-level analyses that individual studies cannot provide. \textcite{Leite_2021} described the result a \enquote{dataquake}. 
Yet the PDB captured the literature only through 2021.  
The thousands of papers published since---containing tens of thousands of additional devices---remain as fragmented and inaccessible as the pre-2021 literature the PDB originally addressed. 
Consequently, the PSC community is currently operating with a multi-year knowledge blackout, unable to systematically assess the aggregate impact of thousands of recent breakthroughs or accurately train data-driven models on the state-of-the-art device architectures.

The bottleneck is human throughput. Extracting device parameters with PDB-level granularity and quality requires deep engagement with each paper. 
In contrast, individual experts can process perhaps dozens of papers annually at this standard. The perovskite literature grows by tens of thousands of papers per year. 
Without infrastructure that converts new publications into structured data automatically, the archive of underutilized knowledge will only expand and limit the progress of the field.
Automation, however, demands reliability:  without matching or exceeding the consistency and quality of human extraction, automatically curated data has limited value.

\begin{figure}[hbt]
    \centering
   \includegraphics[width=\textwidth]{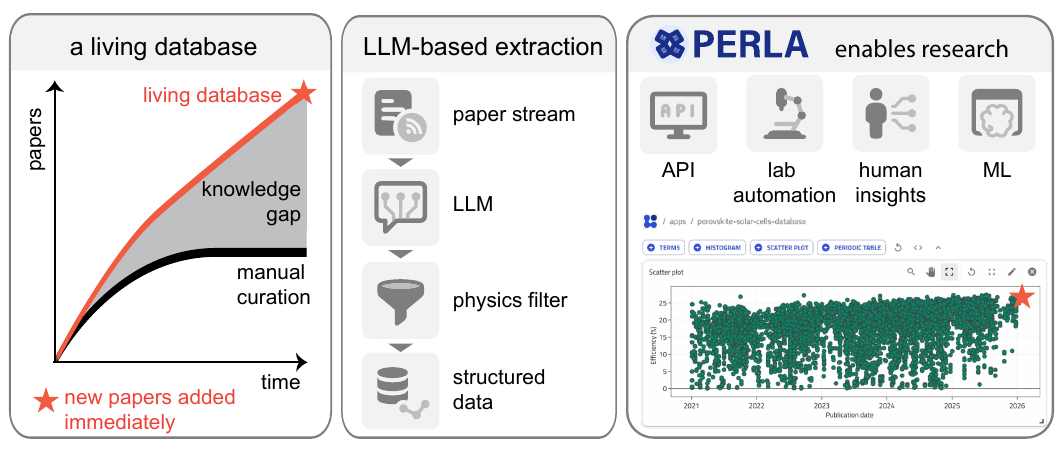}
    \caption{\textbf{Overview of the autonomous living database ecosystem \perla.} While the number of papers published in a field continuously increases, the number of papers in a manually curated database remains constant after curation ends. 
    Living databases continue to grow as the field expands and new results are added automatically (indicated by the red star on the efficiency versus year plot).  
    For this, new scientific literature is ingested via journal RSS feeds (or manual uploads) and processed by a Large Language Model (LLM). Extracted data undergoes physics-based validation filtering before being serialized into a structured JSON format. 
    The \perla ecosystem enables diverse downstream use cases via an application programming interface (API). The API can distribute the continuously updated data to diverse workflows, including retraining machine learning (ML) models to prevent concept drift, facilitating human meta-analysis, and guiding experimental planning in self-driving laboratories. It also enables interactive exploration without programming skills in a graphical user interface.}
    \label{fig:overview}
\end{figure}

Here, we establish the first autonomous, self-updating living database for photovoltaics. By integrating large language models (LLMs) with physics-based constraints, we convert the continuous stream of unstructured literature into a dynamic, structured knowledge resource (the perovskite living archive, \perla) with human-level precision (\Cref{fig:overview}).
The workflow operates autonomously, depositing FAIR-compliant records into an open-access database.
The infrastructure delivers not only self-updating data but self-updating insights. 
With post-2021 literature now flowing continuously into structured form, population-level trends have emerged that static snapshots cannot capture: systematic improvements in voltage losses, field-wide shifts in bandgap targeting, and rapid compositional diversification. 

The implications extend beyond perovskites. Materials science increasingly faces publication rates that outpace organizational capacity, while machine learning---the field's most promising accelerant\autocite{alan_ml_review, Yao_Lum_Johnston_Mejia-Mendoza_Zhou_Wen_Aspuru-Guzik_Sargent_Seh_2022, DeLuna_Wei_Bengio_Aspuru-Guzik_Sargent_2017}---requires fresh, self-updating, and comprehensive training data to avoid concept drift.\autocite{Bayram_Ahmed_Kassler_2022} 
Manual curation, regardless of quality, becomes outdated upon publication. 
This work demonstrates that continuously updating databases can transform how data-driven discovery operates, converting artificial intelligence from an occasional retrospective tool into an integrated research instrument that learns as the field advances.
We release the complete system---extraction pipeline, accumulated data, code, and documentation---as a community resource and as a framework for other rapidly evolving research domains.

%% file: sections/results.tex
\section{Results}
\subsection{Achieving human-level precision through physics-constrained LLMs}

\paragraph{Literature watcher}
The \perla pipeline converts perovskite literature into structured data continuously, without human intervention.
An automated watcher (see \Cref{sec:meth-papersbot} for details) monitors Rich Site Summary (RSS) feeds from relevant journals. Regular expressions (regex) tuned for single-junction perovskite solar cells flag candidate papers. For papers that pass this first screen, we retrieve the full abstracts and apply stricter pattern matching. Review articles and computational studies are excluded. 
Where possible, the system retrieves full texts from open-access papers (e.g., via Unpaywall\autocite{Dhakal_2019})  and ingests the converted PDFs into the LLM-based extraction pipeline (see \Cref{fig:bot_sankey}).

\paragraph{Constrained LLM-based extraction} 
Recent work has shown that LLMs can flexibly extract structured data from unstructured text.\autocite{extraction_review, Polak_Modi_Latosinska_Zhang_Wang_Wang_Hazra_Morgan_2024, polak2024extracting, dagdelen2024structured, Hira_Zaki_Sheth_Mausam_Krishnan_2024, Sayeed_Mohanty_Sparks_2024, Mahjoubi_Venugopal_Manav_AzariJafari_Kirchain_Olivetti_2025, Zheng_Zhang_Borgs_Chayes_Yaghi_2023, Leong_Pablo-García_Wong_Aspuru-Guzik_2025, milica_extraction}
Compared to previous automated extraction approaches\autocite{Swain_Cole_2016, chemdataextractor_2} that have also been explored for perovskites,\autocite{Beard_Cole_2022, Valencia_Liu_Zhang_Bo_Li_Daoud_2025} LLMs excel with greater flexibility and better general understanding of the scientific context\autocite{Mirza2025}: LLMs understand that \enquote{Spiro-OMeTAD} and \enquote{Spiro-MeOTAD} refer to the same molecule, that \enquote{10 min \ce{TiCl4} treatment} and \enquote{\ce{TiCl4} treatment for 10 minutes} describe identical processes, and that an efficiency reported in a table caption belongs to the device described three paragraphs earlier. 
Previous regex-based extractors could not make these connections.

Our extractor targets a device schema aligned with community practice (architecture, stack, processing steps, area, light source, PCE/\(J_\mathrm{SC}\)/\(V_\mathrm{OC}\)/FF, stability, notes) and is a revised version of the schema used in the PDB (see \Cref{sec:schema}). 
We ensure the quality of the extraction with three mechanisms.
First, structured decoding enforces valid field names and data types—the model cannot invent new schema fields. Second, unit normalization converts all quantities to canonical forms and flags dimensional mismatches. Third, physics validation eliminates Shockley-Queisser violations and other unphysical combinations. Ions are automatically normalized into a dedicated schema (see \Cref{sec:nomad_app}).\autocite{maqsood_towards_2025}

To establish the reliability necessary for autonomous scientific data gathering, we evaluated our system against a human-annotated ground truth.
For this, we constructed a ground truth dataset informed by manual annotation by 13 domain experts---faculty, postdocs, and PhD students with active publication records in perovskite research. Full consensus among all annotators occurred in only 13\%. 
This indicates that data from manual extraction campaigns might be impacted by large variance due to human variability. This source of variance is removed in automated extraction workflows such as the one in \perla (see \Cref{fig:comparison_with_human_performance}).

\begin{figure}[htb]
    \centering
    \includegraphics[width=\linewidth]{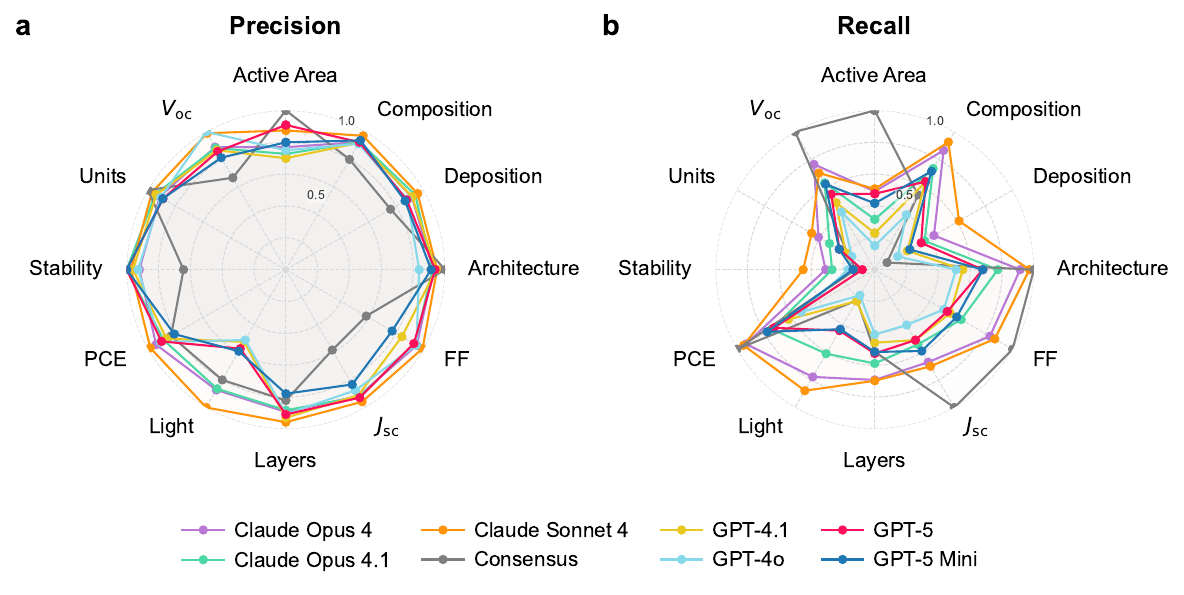}
    \caption{\textbf{Performance of the extraction pipeline.} All metrics shown are micro averages, i.e., averaged over all cells in the ground truth. Lines show different models. \enquote{Consensus} represents an aggregation of human labelers.  
     \textbf{a.} Precision indicates how many of the extracted entries were correct, i.e., matching the ground truth. For numeric fields, this involves checking if a value is in a given tolerance range. A PCE precision of 0.9, for instance, means that \SI{90}{\percent} of extracted PCE values agree within the tolerance with the ground truth. In the development of the system, we prioritized precision.  \textbf{b.} Recall measures how many of the entries that are in the ground truth have been extracted. A lower recall means that fewer entries have been extracted. Typically, there is a tradeoff between precision and recall and it is difficult to optimize both at the same time. }
    \label{fig:performance}
\end{figure}

We evaluated multiple state-of-the-art models against 20 papers (101 solar cells) where the ground truth was derived using arbitration to resolve discrepancies between human annotations (\Cref{fig:performance}). Our evaluation uses the Munkres algorithm\autocite{kuhn1955hungarian, munkres1957algorithms} to pair extracted devices with ground truth based on composition, architecture, and processing similarity, then calculates precision and recall on matched pairs. For semantic fields like material names, we employ an LLM-as-judge\autocite{li2024llmsasjudges} (GPT-4o, validated at 97\% accuracy on curated examples, see \Cref{sec:extraction_pipeline}) to assess equivalence.

Claude Sonnet 4, Claude Opus 4.1, and GPT-4.1 achieved precision exceeding 0.9 across all field categories, meaning that the extracted values agree (within tolerance) with the values in the ground truth in \SI{90}{\percent} of the cases. 
GPT-4o, GPT-5, and GPT-5 Mini showed lower performance. Based on its good performance, we used Claude Sonnet 4 for the final pipeline.
We optimized for precision over recall, prioritizing accuracy over completeness.
Comparing model performance to inter-annotator agreement reveals that automated extraction errors are comparable to or smaller than human inconsistencies (see \Cref{fig:comparison_with_human_performance} for comparison with human annotators). 
Overall, the system achieves human-level reliability with a critical advantage---it applies identical extraction logic uniformly, eliminating annotator-to-annotator variance that plagues manual curation.
To ensure a high quality of the extraction results, we apply additional filters for hallucination and physical consistency, which can make the automatically generated database more consistent than the manually created one (see \Cref{fig:physics_consistency_check}).

\paragraph{Publishing to a FAIR backbone} 
Validated entries deposit automatically into NOMAD, a research data management platform for materials science.\autocite{Scheidgen2023}
Each entry carries DOI back-links, ensuring provenance. The infrastructure provides programmatic API access and interactive dashboards.  
In total, the LLM stream contributes 2582 publications and 7816 devices beyond the manual PDB baseline.

\subsection{Materials insights from a living resource}
With post-2021 literature flowing into structured form automatically, we can now observe the evolution of perovskite solar cells in near real time.
In the following, we focus on three emerging patterns revealed by the continuously updated dataset: sustained performance improvements driven by voltage loss reduction, rapid compositional diversification within iodide perovskites, and an ongoing architectural shift toward inverted device structures with advanced interfacial layers.

\paragraph{Continued voltage loss reduction}

\begin{figure}[!htb]
    \centering
    \includegraphics[width=.7\textwidth]{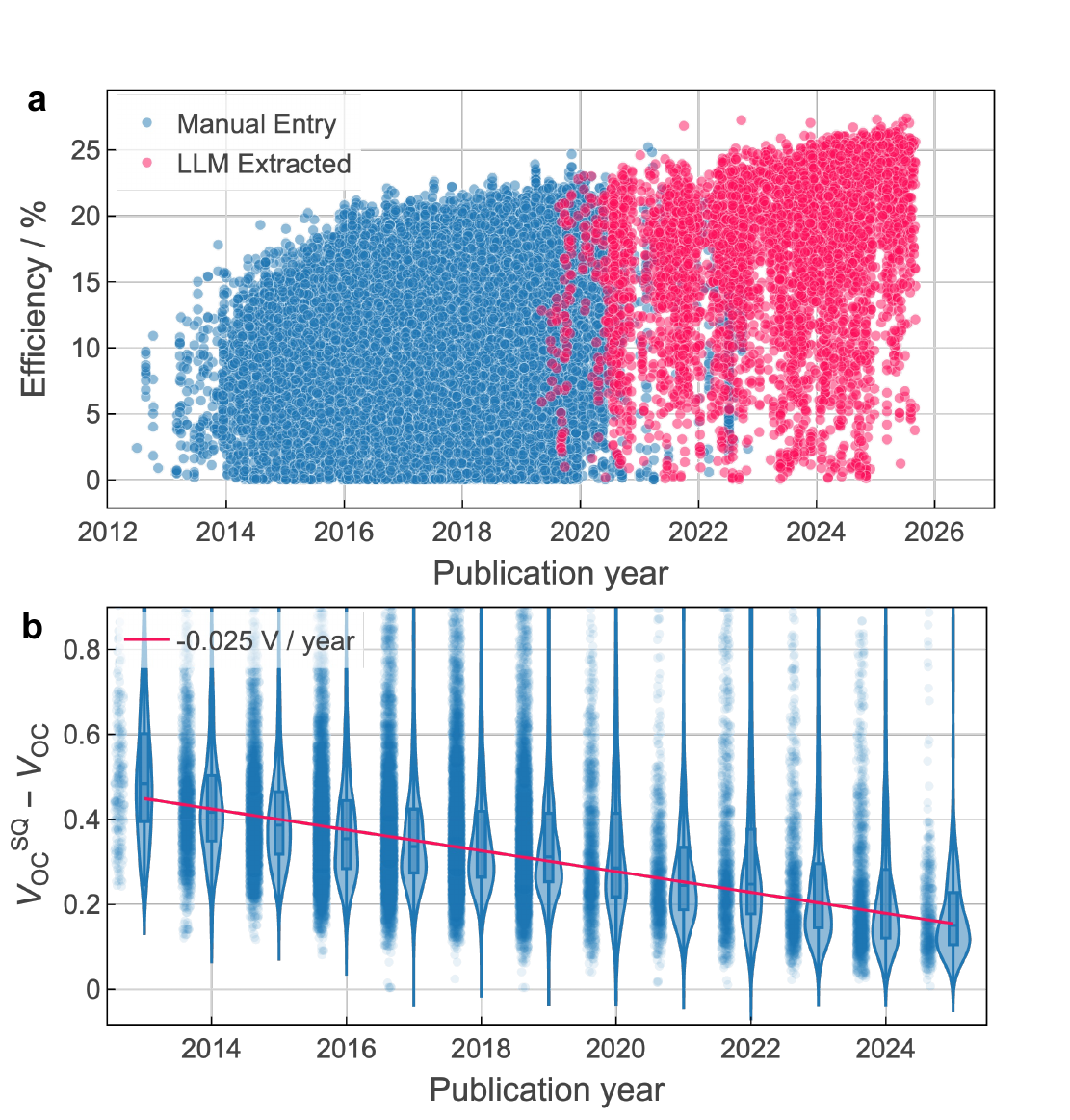}
    \caption{\textbf{\perla reveals the continued improvement of perovskite solar cells.} \textbf{a} Comparison of reported perovskite solar cell efficiencies over time between the curated Perovskite Solar Cell Database (blue) and values automatically extracted from the literature using an LLM (purple). \textbf{b} Yearly distribution of the voltage loss, defined as the difference between the Shockley--Queisser (SQ) limit $V_\text{OC}^\text{SQ}$ and the reported open-circuit voltage $V_\text{OC}$. The red line represents a linear fit ($R^2 = 0.97$), showing a steady reduction in voltage loss of approximately \SI{25}{\milli\volt} per year.}
    \label{fig:pce-evolution}
\end{figure}

We observe a sustained increase in reported power conversion efficiencies (PCEs) over time (\Cref{fig:pce-evolution}\textbf{a}).
To disentangle performance gains from compositional and optical effects, we analyze the voltage deficit, defined as the difference between the Shockley-Queisser\autocite{Shockley_Queisser_1961} open-circuit voltage limit $V_\mathrm{OC}^\mathrm{SQ}$ and the experimentally reported $V_\mathrm{OC}$.
This quantity provides a normalized, physically meaningful measure of recombination losses that allows comparisons across devices with different bandgaps.

Despite substantial device-to-device variability within each year, the median voltage deficit exhibits a strikingly linear decrease over time (\Cref{fig:pce-evolution}\textbf{b}).
A linear fit yields a reduction rate of approximately \SI{25}{\milli\volt} per year ($R^2 = 0.97$), indicating steady progress in mitigating recombination losses.
Notably, this trend shows no sign of saturation to date.
While reaching the radiative limit---corresponding to near-unity external quantum efficiency of luminescence---is only a theoretical bound and unlikely to be fully realized experimentally, the persistence and linearity of the observed trend underscore the continued effectiveness of loss-reduction strategies in the field.

\paragraph{Compositional diversification}

To analyze compositional trends in a consistent and interpretable manner, we restrict the following analysis to iodide-based halide perovskites and a defined set of commonly used A-site cations.
This simplification is enabled by the extraction pipeline, which automatically normalizes reported compositions against a curated ions database \autocite{maqsood_towards_2025} and formalizes diverse textual descriptions into standardized halide perovskite formulas.
This approach allows robust population-level statistics while avoiding ambiguities arising from inconsistent nomenclature.

\begin{figure}[!htb]
    \centering
    \includegraphics[width=1.0\textwidth]{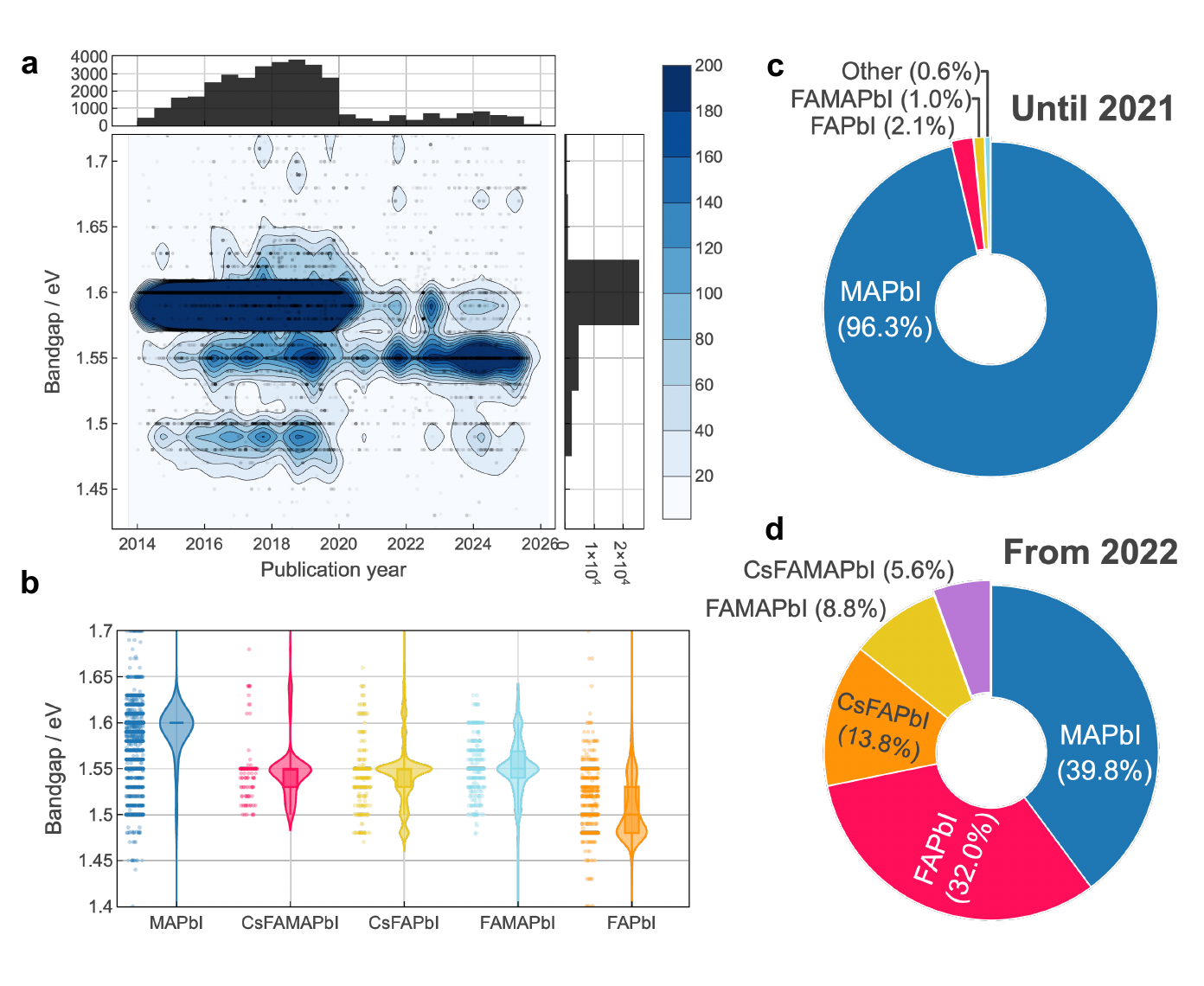}
    \caption{\textbf{Evolution of bandgap values and perovskite compositions over time.}
    \textbf{a} Kernel density estimation (KDE) of reported bandgap values as a function of publication year, with a histogram of publication counts. The data shown are limited to bandgap values between 1.2 and \SI{2.2}{\electronvolt} ($n = 36\,788$).
    \textbf{b} Distribution of bandgap values represented as violin plots for pure iodide perovskites in the selected compositions ($n = 27\,593$).
    \textbf{c} Relative frequency of reported short-form compositions until 2021, dominated by MAPbI (\SI{96.3}{\percent}), based on 26\,342 entries.
    \textbf{d} Relative frequency of compositions from 2022 onwards, showing diversification with increased prevalence of FAPbI (\SI{32.0}{\percent}), FAMAPbI (\SI{8.8}{\percent}), and CsFAPbI (\SI{13.8}{\percent}) ($n = 1\,777$). Panels \textbf{b--d} are based on a subset of pure iodide perovskites for which both the absorber composition and an explicit bandgap value could be retrieved from the literature. The analysis was performed for selected compositions (MAPbI, CsFAMAPbI, CsFAPbI, FAMAPbI, FAPbI).}
    \label{fig:bandgap_compositions}
\end{figure}

Within this normalized compositional space, the dataset documents a pronounced transition in absorber materials when we filtered entries in which we have both the composition of the absorber and the bandgap value.
Before 2021, methylammonium lead iodide (\ce{MAPbI3}) dominated the literature, accounting for more than \SI{96}{\percent} of reported devices.
In contrast, post-2021 publications show a marked diversification:  Formamidinium lead iodide (\ce{FAPbI3}) represents approximately \SI{32}{\percent} of devices, while mixed-cation compositions such as \ce{FAMAPbI3} and \ce{CsFAPbI3} account for \SI{8.8}{\percent} and \SI{13.8}{\percent}, respectively.
Over the same period, the share of \ce{MAPbI3} devices decreased to \SI{39.8}{\percent} (see \Cref{fig:bandgap_compositions}).

This compositional shift is accompanied by a systematic change in reported bandgap values.
Kernel density estimates and violin plots in \Cref{fig:bandgap_compositions} show that the modal bandgap moves from approximately \SI{1.6}{\electronvolt}, characteristic of \ce{MAPbI3}, toward the \SIrange{1.50}{1.55}{\electronvolt} range associated with FA-rich compositions.
This shift brings absorber bandgaps closer to the optimum for single-junction solar cells, increasing the maximum attainable Shockley-Queisser efficiency.
The temporal coincidence of this bandgap shift with the observed efficiency gains suggests that compositional optimization complements ongoing reductions in recombination losses.

\paragraph{Architectural transformation}

\begin{figure}
    \centering
    \includegraphics[width=\textwidth]{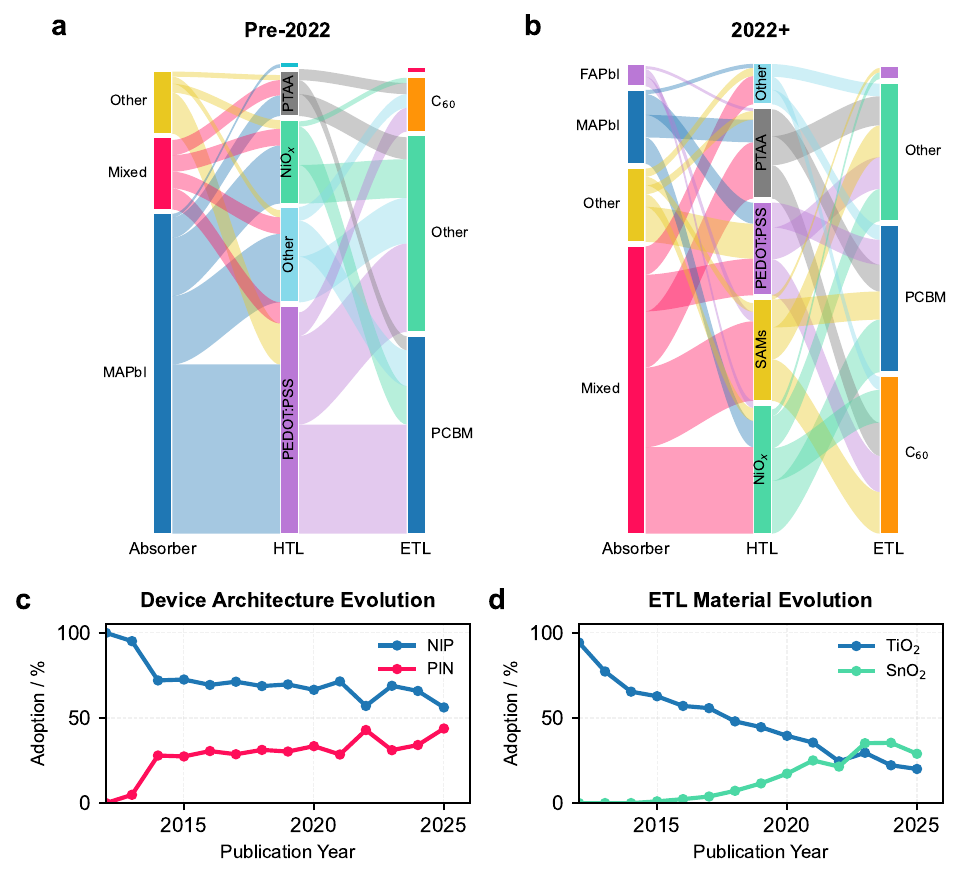}
\caption{\textbf{Temporal evolution of materials and device architectures.} 
\textbf{a, b,} Sankey diagrams illustrating the shifting landscape of material combinations for \textit{p-i-n} architectures for the \textbf{a,} pre-2022 and \textbf{b,} 2022+ eras. The flow width represents the relative frequency of material pairings between the perovskite absorber, hole transport layer (HTL), and electron transport layer (ETL). Notable trends include the transition from MAPbI toward mixed-cation absorbers and the emergence of self-assembled monolayers (SAMs) in the HTL position. The evolution of \textit{n-i-p} devices is shown in \Cref{fig:nip_evolution}.
\textbf{c,} Adoption rates of device architectures from 2013 to 2025. The data highlights the historical dominance of the \textit{n-i-p} (conventional) structure and the steady rise in \textit{p-i-n} (inverted) configurations, which reached over 40\% adoption by 2025. 
\textbf{d,} Evolution of ETL material preferences, showing the displacement of \ce{TiO2} by \ce{SnO2} as the primary electron-selective contact. }
    \label{fig:architectural_transformation}
\end{figure}

Device architecture changed in parallel with composition (see \Cref{fig:architectural_transformation}). 
\textit{n-i-p} structures dominated in 2010. \textit{p-i-n} (inverted) architectures increased to over \SI{43}{\percent} of devices by 2025.

Hole transport materials evolved with device architecture (\Cref{fig:architectural_transformation}). Spiro-OMeTAD dominated \textit{n-i-p} devices (\Cref{fig:nip_evolution}). In \textit{p-i-n} structures, PEDOT:PSS and PTAA gained use initially. 
Strikingly, the data uncovers an explosive adoption of self-assembled monolayers (SAMs)\autocite{sam_adv_ener_mat_2018, al-ashouri_monolithic_2020, Tang_Shen_Shen_Yan_Wang_Han_Han_2024} as hole-transport materials, a trend almost entirely absent from pre-2021 databases.

\begin{figure}[htb]
    \centering
    \includegraphics[width=\textwidth]{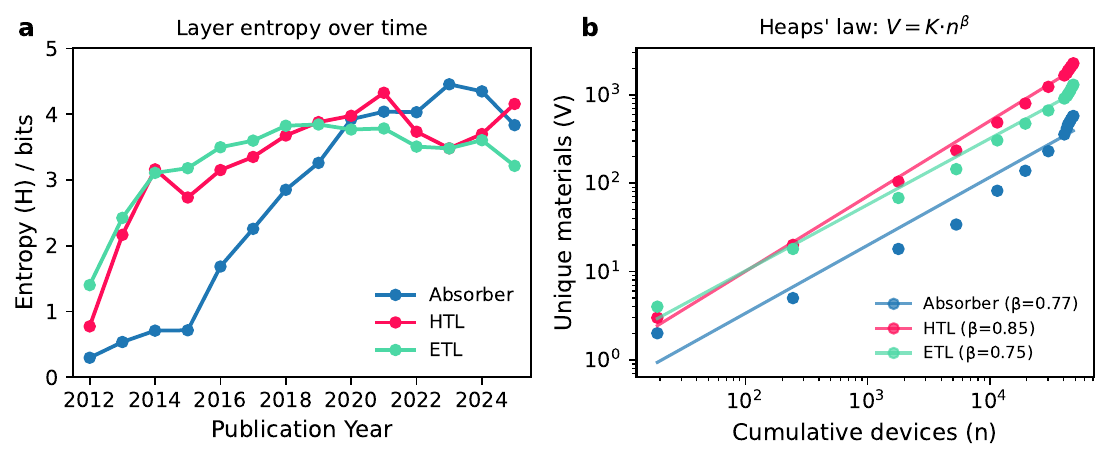}
    \caption{\textbf{Stack diversity over time.} \textbf{a} Annual Shannon entropy $H$ for absorber, hole-transport layer (HTL), and electron-transport layer (ETL) categories; larger $H$ denotes more even use of multiple materials within a layer. An entropy of 0 means that only one material would be used; a higher entropy means that usage is spread over materials. \textbf{b} Fit to heaps' law. The material vocabulary shows sublinear growth, indicating diminishing discovery rate per paper, which is a signature of field maturation. }
    \label{fig:entropy_diversity}
\end{figure}

Electron transport materials transformed, too. While the initial focus in \textit{p-i-n} architectures was on PCBM materials, \ce{C_{60}} gained prominence. Overall, we find that the material flows are more fragmented (more flows with meaningful width in \Cref{fig:architectural_transformation}) --- for each material, there tend to be more materials with which it is combined. 
\Cref{fig:entropy_diversity} provides a quantitative assessment of this maturation through the lens of information theory and scaling laws. First, we characterize the diversity of material utilization using Shannon entropy ($H$), where $H=0$ represents the absolute dominance of a single material and increasing values reflect a more uniform distribution across the discovered chemical space. We observe a period of rapid \enquote{stochastic experimentation} in the field's infancy (2012–2016), evidenced by a sharp increase in entropy across all device layers. Notably, experimentation in transport layers preceeded the experimentation in the absorber.
The subsequent stabilization suggests that the field has effectively converged toward a consensus set of \enquote{champion} absorber compositions, shifting the focus from the search for novel bulk materials to the fine-tuning of established high-performance systems.
This interpretation of a maturing field is further corroborated by an analysis of the material discovery rates through Heaps' Law ($V = K \cdot n^\beta$), which relates the number of unique materials ($V$) to the cumulative number of reported devices ($n$). As shown in \Cref{fig:entropy_diversity}\textbf{b}, the scaling exponent $\beta$ serves as a proxy for the \enquote{innovation velocity} of each layer. The absorber layer exhibits a sub-linear scaling ($\beta = 0.77$), confirming that new reports increasingly rely on a known library of materials rather than novel discovery. The hole-transport layer (HTL) maintains the highest exponent ($\beta = 0.85$), highlighting that interface engineering remains the primary frontier of innovation. Together, these metrics demonstrate a fundamental shift in the research landscape: the community has transitioned from a broad exploration of perovskite chemistries to a specialized refinement of device architectures, where each subsequent thousand devices yields a predictable, but diminishing, volume of truly novel material combinations.

This evolution of the field also matters for the training of machine learning models. They promise to be a powerful research accelerant. But with manual extraction, they are bottlenecked by only being trained on outdated data. 
If we train models on matched training sets of historic and mixed persists at all training set sizes, we find that compositional coverage, not merely data quantity, determines model generalization (\Cref{sec:ml-case-study}). This underscores the importance of continuously updated databases for machine learning applications in materials science.\\


\noindent \perla is a foundation for data-driven research on perovskites. 
We provide the database, extraction pipeline, and analysis tools as open resources. The data is hosted on NOMAD with programmatic application programming interface (API) access and interactive dashboards. Jupyter notebooks\autocite{kluyver2016jupyter} for custom analyses and the automated monitoring bot are maintained and updated as new publications appear. \Cref{sec:nomad_app} gives examples of analyses, e.g., about Sn-based perovskite solar cells or stability trends, that are made possible with this interactive interface. 
 
\perla can be easily extended. In future work, we will leverage this to address some of the current limitations. Extraction from supplementary materials and figures is not implemented. We use Claude Sonnet 4 for extraction; work is underway to fine-tune open language models to reduce reliance on proprietary systems.\autocite{alampara2025general} Extraction is also limited by the limited access to the full text of papers. We expect this to improve with the increasing success of open-access publications and collaborations with publishers.

%% file: sections/conclusions.tex
\section{Conclusions}
The acceleration of scientific discovery has created a paradox: the community generates knowledge faster than it can be synthesized. In fast-moving fields like photovoltaics, manual curation---historically the gold standard---has collapsed under the weight of exponential publication rates. This has created a widening knowledge gap where data-driven discovery is ironically stifled by the very volume of research it seeks to analyze---forced to rely on static datasets that are obsolete upon release.

We resolve this bottleneck by establishing the first autonomous, self-updating living database for perovskite solar cells. By coupling LLMs with physics-aware validation, we demonstrate that an automated system can extract complex device parameters with human-parity precision ($>90\%$) while eliminating annotator variance. The result is a dynamic, real-time view of the field that immediately revealed post-2021 trends---such as the rapid dominance of inverted architectures and self-assembled monolayers---that remained invisible to static curation efforts.

The implications extend beyond perovskites. We have demonstrated that the \enquote{data velocity} problem in materials science is solvable. By integrating LLMs with mature research data management infrastructure, we convert the collective knowledge of the scientific literature from a static archive into a continuously learning knowledge resource. We release this entire infrastructure as an open-source blueprint, offering a scalable path for any scientific domain to operate at the true pace of discovery.

%% file: sections/methods.tex
\section{Methods}

\paragraph{Literature watcher}\label{sec:meth-papersbot}
We built an automated system to identify new perovskite solar cell publications. The system monitors RSS feeds from major journals using regular expressions tuned for single-junction devices. Initial filtering uses permissive patterns to avoid false negatives from truncated RSS summaries. Candidate papers are validated through abstract retrieval from multiple sources (CrossRef, OpenAlex,\autocite{priem2022openalex0} Semantic Scholar\autocite{kinney2023semantic}, and PubMed\autocite{sayers2025database}) followed by strict pattern matching. Further filtering is done to exclude review articles and simulation studies (details in \Cref{sec:filtering_pipeline}). The open-access full texts are retrieved via Unpaywall and OpenAlex when available and ingested into the LLM-based extraction pipeline. The extracted cells are then normalized and validated before being uploaded to NOMAD.
Over a representative 106-day period, the system parsed 91387 papers, identifying 311 relevant publications, of which 75 were open-access and proceeded to extraction (\Cref{fig:bot_sankey}).

\paragraph{Bulk extraction}
To capture post-2021 literature systematically, we performed a large-scale search using the Scopus API.  We filtered for relevant journals based on CiteScore rankings, feedback from our human labelers and excluded journals focused on reviews, theory, computation, or other fields of chemistry and materials science. Complete search strategy is described in  \Cref{appendix:search}.

\paragraph{LLM-based data extraction}

We use the structured output capabilities of modern LLMs through function-calling endpoints, which constrain model generation to produce valid JSON conforming to our device schema. This eliminates format parsing errors and ensures all outputs match expected data structures. 
We optimized the extraction pipeline using a separate validation set, distinct from the held-out test set used for final evaluation. The validation set comprised 10 papers with expert annotations. We iteratively refined extraction prompts and evaluated multiple state-of-the-art models (GPT-4.1, GPT-4o, GPT-5, GPT-5 Mini, Claude Opus 4, Claude Opus 4.1, Claude Sonnet 4) on this validation set to select the optimal model and prompt configuration.

\paragraph{Evaluation}
We manually annotated 20 papers as the test set and 10 papers as the validation set. The test set was bootstrapped based on a campaign in which thirteen domain experts (faculty, postdocs, PhD students with active publication records in perovskite research) manually annotated papers (see \Cref{sec:manual_annotation}). The final ground truth was determined by arbitration in the case of disagreement between human annotators.  
Evaluation uses a three-stage procedure. First, all numerical values undergo unit conversion to canonical forms using the pint library.\autocite{pint} The system recursively traverses nested JSON structures, identifies fields with numerical values and units, determines physical dimensionality, and converts to standard units.
Second, the Munkres algorithm\autocite{kuhn1955hungarian, munkres1957algorithms} finds optimal pairings between extracted and ground truth devices.
Third, semantic fields like material names are evaluated using an LLM-as-judge (GPT-4o, validated at 97\% accuracy on 43 curated cases). The evaluation pipeline is described in detail in \Cref{sec:extraction_pipeline}.

%% file: sections/appendix.tex
\section{Data model} \label{sec:schema}
A simplified version of our data model is shown in the following listing. The full data model can be found on GitHub. 
During data upload and processing in NOMAD, an additional mapping to the former data models of the Perovskite Database\autocite{perovskite_db} is also created and stored to facilitate aggregation with legacy data.

\begin{mintedbox}{python}
class PerovskiteSolarCells(BaseModel):
    cells: List[PerovskiteSolarCell]

class PerovskiteSolarCell(BaseModel):
    pce: PCE                              # {value: float, unit: "
    jsc: JSC                              # {value: float, unit: "mA cm^-2"}
    voc: VOC                              # {value: float, unit: "V"|"mV"}
    ff: FF                                # {value: float}
    device_architecture: Literal["pin", "nip", ...]
    active_area: ActiveArea
    light_source: LightSource
    stability: Stability
    perovskite_composition: PerovskiteComposition
    layers: List[Layer]

class PerovskiteComposition(BaseModel):
    formula: str                          # e.g., "FA0.85MA0.15PbI2.55Br0.45"
    dimensionality: Literal["0D", "1D", "2D", "3D", "2D/3D"]
    sample_type: Literal["Polycrystalline film", "Single crystal", ...]
    bandgap: Bandgap
    a_ions, b_ions, x_ions: List[Ion]     # A/B/X site ions

class Ion(BaseModel):
    abbreviation: str                     # "FA", "MA", "Cs"
    common_name: str                      # "Formamidinium"
    molecular_formula: str                # "CH6N+"
    coefficient: str                      # "0.85", "1-x"

class Layer(BaseModel):
    name: str                             # "TiO2", "Spiro-MeOTAD"
    functionality: Literal["ETL", "HTL", "Absorber", "Contact", "Substrate"]
    thickness: Thickness
    deposition: List[ProcessingStep]

class ProcessingStep(BaseModel):
    method: str                           # "Spin-coating", "Evaporation"
    atmosphere: Literal["N2", "Air", "Vacuum", ...]
    temperature: Temperature
    duration: Time
    solution: ReactionSolution

class Stability(BaseModel):
    time: Time
    PCE_T80: Time                         # time to 80
    humidity: Humidity
    temperature: Temperature
\end{mintedbox}

\section{Literature watcher} \label{sec:apx-papersbot}
We built the literature watcher bot based on PapersBot\autocite{PapersBot} to monitor journal RSS feeds for new perovskite solar cell papers and extract data from them to NOMAD. For an initial coarse filtering, we use a permissive simple regular expression as often the summary obtained from the feeds is either truncated or simplified, and thus is missing crucial keywords and identifying information. 
Using a strict regex against these results in a high false negative rate. For the coarse matched papers, we try to obtain abstracts from Crossref with OpenAlex, Semantic Scholar, and PubMed as cascading fallbacks. The abstracts and the summaries are then matched against the strict regex to get the candidate set of papers that are then further filtered using the pipeline detailed in \Cref{sec:filtering_pipeline}.

\begin{figure}[!h]
    \centering
    \includegraphics[width=1\linewidth]{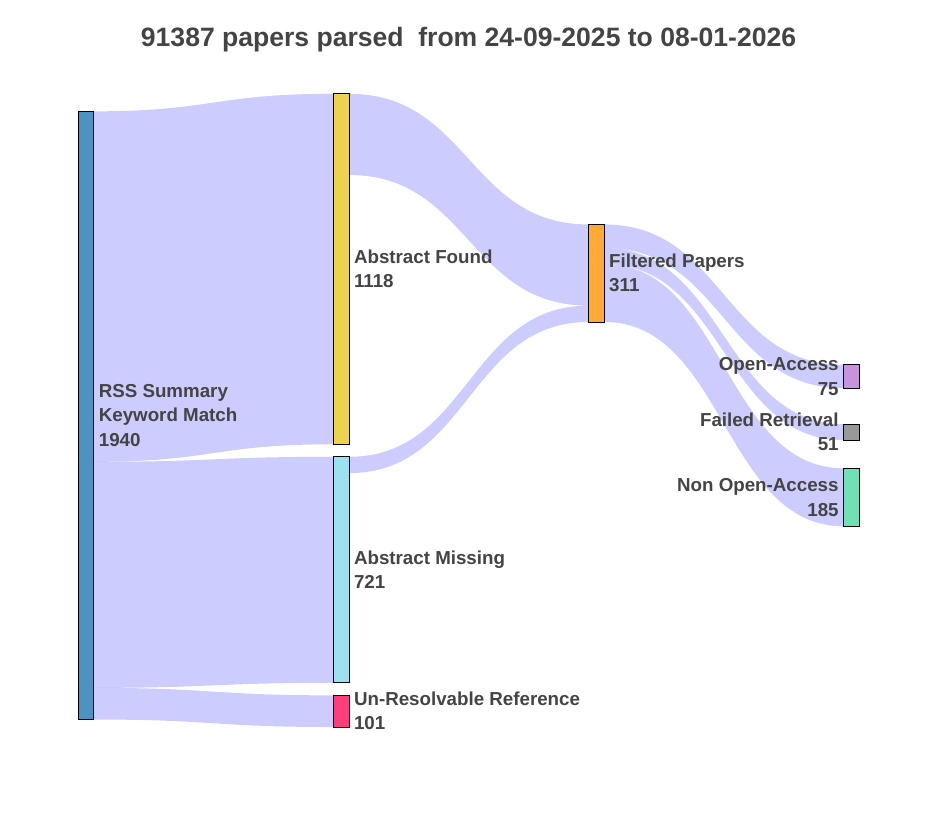}
    \caption{\textbf{The multi-stage filtering pipeline for identifying new perovskite solar cell papers.}  Over a 106-day period (24-09-2025 to 08-01-2026), 91387 papers were parsed from Journal RSS feeds. An initial match against RSS summaries identified 1940 candidates. Subsequent steps remove papers with unresolvable DOIs ($n = 101$), and failing a secondary strict match ($n = 1348$). Further filtering is done to exclude theoretical, computational and review works ($n = 180$), yielding a final set of 311 relevant papers of which 75 were open-access papers.}
    \label{fig:bot_sankey}
\end{figure}

For the period 24-09-2025 to 08-01-2026, 91387 relying only on the RSS summaries for the strict matching, we get an initial set of 161 papers, of which 101 papers pass further filtering, yielding a final set of 13 open-access papers. Retrieving the abstract before applying stricter regex matching increases the number of relevant papers to 330, from which, after further filtering, we get an additional 62 open source papers to extract. \Cref{fig:bot_sankey} shows a summary of the pipeline for the same period. 


As papers are often not immediately indexed or available as open access after their publication, the pipeline retries the retrieval of abstracts and checking open access status with an increasing time delayed retry schedule.

\section{Literature search} \label{appendix:search}
To construct a high-quality corpus of peer-reviewed articles relevant to perovskite solar cells, we performed a systematic literature search using the Scopus API. 
Our query was designed to retrieve journal articles containing the terms \texttt{solar cell}, \texttt{pce}, and \texttt{perovskite}, published between 2021 and 2025. The query string used was:

\begin{quote}
\texttt{SRCTYPE(j) AND solar cell AND pce AND perovskite}
\end{quote}

Separate API calls were made for each publication year within this range, and results were paginated in batches of 25 entries. 
Only articles indexed as journal publications (\texttt{SRCTYPE(j)}) were considered.

To ensure the inclusion of only relevant literature, we preselected journals based on a curated list of journals.  

After retrieving the initial list of candidate articles, we extracted DOIs from the Scopus API response and resolved them via the CrossRef API, with OpenAlex, Semantic Scholar, and PubMed used as fallback sources, to obtain associated journal names and publishers. 
Each article was subjected to additional filtering based on journal metadata and title heuristics removing articles focussed on simulations or review articles and perspectives.

The full filtering and selection process was automated using a custom Python pipeline.

The final list of DOIs was manually downloaded for further extraction through our data extraction pipeline. 

\section{Filtering pipeline} \label{sec:filtering_pipeline}

To ensure the integrity and relevance of the dataset, we implemented a multi-stage automated filtering pipeline to select articles for data extraction. The filtering process operates on DOIs and retrieves article metadata and abstracts using a prioritized sequence of external sources. Metadata is queried first from the Crossref API, with automatic fallback to OpenAlex, Semantic Scholar, and PubMed when required, ensuring robust coverage in cases of missing or incomplete records. Full-text analysis is subsequently applied where necessary.

\subsection{Metadata retrieval and text normalization}
For each candidate DOI, metadata (title, abstract, journal name, and publisher) was fetched from our sources. To standardize the input for content analysis, the text source was dynamically selected:
\begin{itemize}
    \item If the abstract contained fewer than 100 words or was unavailable, we utilized the first 5\% of the article's full PDF text  as the screening corpus.
    \item Otherwise, the combined title and abstract served as the screening corpus.
\end{itemize}
Text normalization was applied to handle variations in journal names (e.g., removal of conjunctions and ampersands).

\subsection{DOI filtering criteria}
A candidate article was included in the final dataset only if it passed four sequential filters.

\subsubsection{Journal scope and relevance}
We restricted the dataset to sources likely to contain high-quality experimental data on PSCs. The journal filtering logic applied two conditions:
\begin{itemize}
    \item \textbf{Exclusion:} Journals with titles containing keywords unrelated to experimental device fabrication (e.g., \enquote{toxicology}, \enquote{catalysis}, \enquote{theory}, \enquote{computation}, \enquote{reviews}) were automatically excluded.
    \item \textbf{Inclusion list:} The journal name was cross-referenced against a manually curated whitelist of relevant publications. Fuzzy matching was employed to account for minor naming variations.
\end{itemize}

\subsubsection{Topic classification (solar vs. non-solar)}
To distinguish PSC research from other perovskite applications (e.g., LEDs, photodetectors, or sensors), we implemented a frequency-based keyword voting system. We defined two keyword sets:
\begin{itemize}
    \item \textbf{Solar keywords ($S$):} e.g., \enquote{solar cell}, \enquote{photovoltaic}, \enquote{PCE}, \enquote{PV}.
    \item \textbf{Non-solar keywords ($O$):} e.g., \enquote{LED},  \enquote{battery}, \enquote{laser}, \enquote{transistor}, \enquote{catalyst}.
\end{itemize}
An article was deemed relevant only if the count of solar keywords ($N_{S}$) satisfied two conditions:
\begin{equation}
    (N_{S} > 0) \quad \land \quad (N_{S} \geq N_{O})
\end{equation}
This ensured that the primary focus of the text was photovoltaic applications.

\subsubsection{Exclusion of theoretical and computational studies}
As the objective was to extract experimental device parameters, we excluded purely theoretical works. We scanned the text corpus (i.e., title and abstract or first \SI{10}{\percent} of full text) for an extensive list of computational keywords, including but not limited to:
\begin{itemize}
    \item \textbf{Simulation methods:} \enquote{DFT}, \enquote{molecular dynamics}, \enquote{VASP}, \enquote{SCAPS}, \enquote{first-principles}.
    \item \textbf{Data-driven methods:} \enquote{machine learning}, \enquote{neural network}, \enquote{deep learning}.
\end{itemize}
Any occurrence of these terms resulted in the immediate rejection of the DOI. While this might lead to false-positives, it ensures that the corpus remains fully experimental. Future work might enhance the filter using LLM-based classifiers.

\subsubsection{Exclusion of secondary literature}
To prevent the extraction of aggregated or duplicate data, review articles and perspectives were filtered out. The pipeline detected these via specific title and abstract markers, such as \enquote{Review}, \enquote{Perspective}, \enquote{Progress in}, \enquote{State of the art}, or \enquote{Recent advances}.

\section{Manual labeling and ground truth dataset construction} \label{sec:manual_annotation}

\subsection{Expert recruitment and dataset selection}

To construct a high-quality ground truth dataset for evaluating automated extraction performance, a panel of 13 domain experts was recruited. 
The panel comprised faculty members, postdoctoral researchers, and PhD students with active publication records in perovskite solar cell research.

A total of 68 representative publications were selected from high-impact journals, including \textit{Nature Energy}, \textit{Science}, \textit{Advanced Materials}, and \textit{ACS Energy Letters}, covering the years 2018--2023. 
Selection criteria emphasized diversity in device architectures, fabrication methodologies, and reported performance metrics.

\subsection{Manual labeling protocol}

\begin{figure}[!h]
    \centering
    \includegraphics[width=1\linewidth]{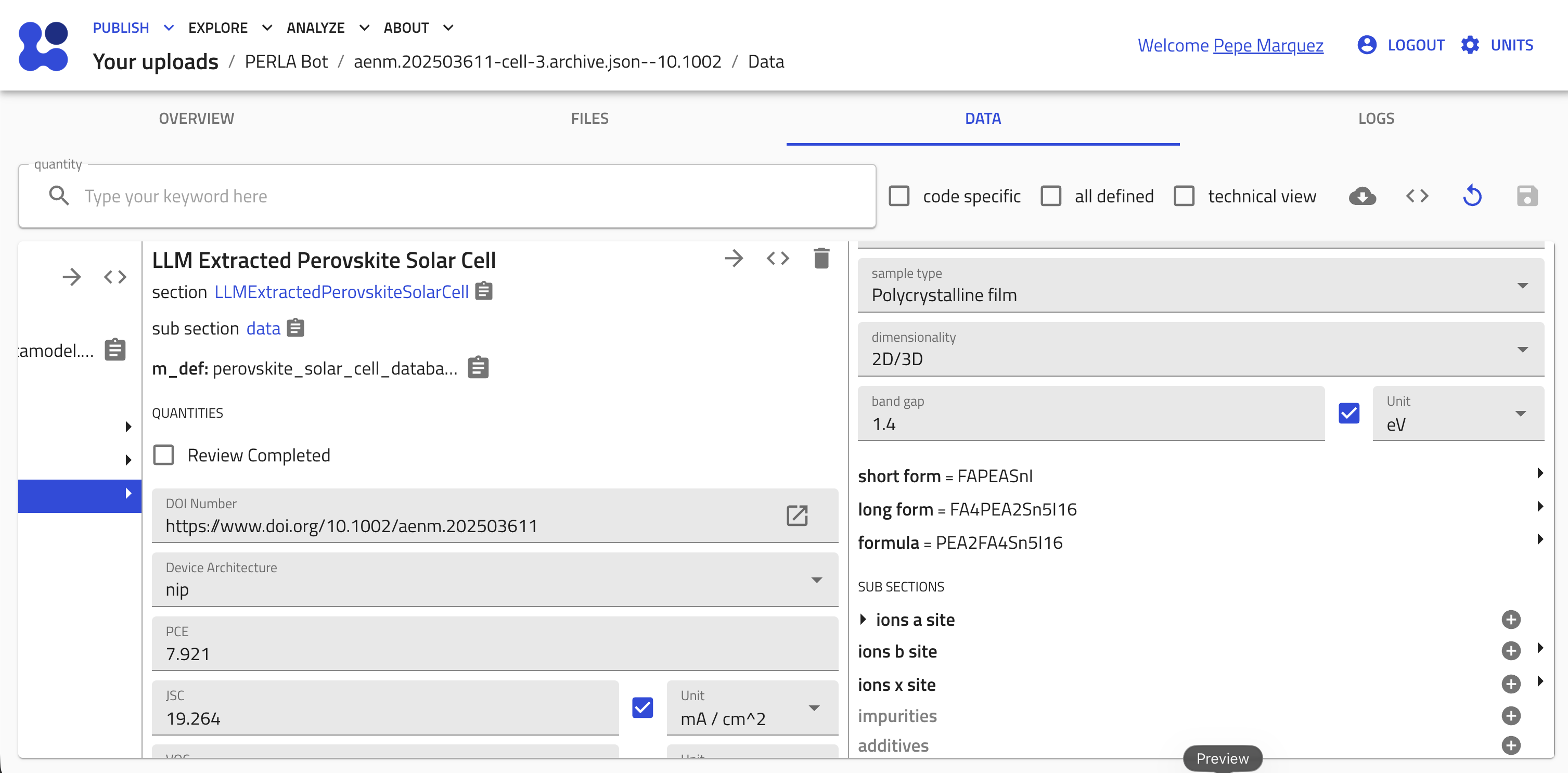}
    \caption{\textbf{NOMAD manual labeling.} NOMAD interface used by human labelers to fill perovskite solar cell data from the assigned literature. NOMAD enables structured labeling of device-level performance metrics and layer-resolved properties within a standardized data schema.}
    \label{fig:manual-labeling-nomad}
\end{figure}

Experts were assigned subsets of publications and asked to verify and correct structured device-level information using a pre-filled annotation interface. To streamline the labeling process and reduce annotator burden, initial extractions were generated using an early version of our automated pipeline and imported into the NOMAD platform. 
Experts used the NOMAD GUI to review, complete, and amend the pre-populated entries, to capture as much relevant device information as present in each paper (see \Cref{fig:manual-labeling-nomad}). 
To ensure consistency across annotators, a shared set of written instructions and labeling guidelines was also provided.

Despite these measures, significant inter-annotator variability was observed. Disagreements frequently arose from inconsistencies in how devices were described in the text, particularly when multiple devices were evaluated within a single study. In several cases, different performance metrics referenced within the same paragraph corresponded to different devices from the experimental pool, without being explicitly distinguished. As a result, annotators diverged in determining which textual performance values belonged to the same device.

\begin{figure}[!h]
    \centering
    \includegraphics[width=1\linewidth]{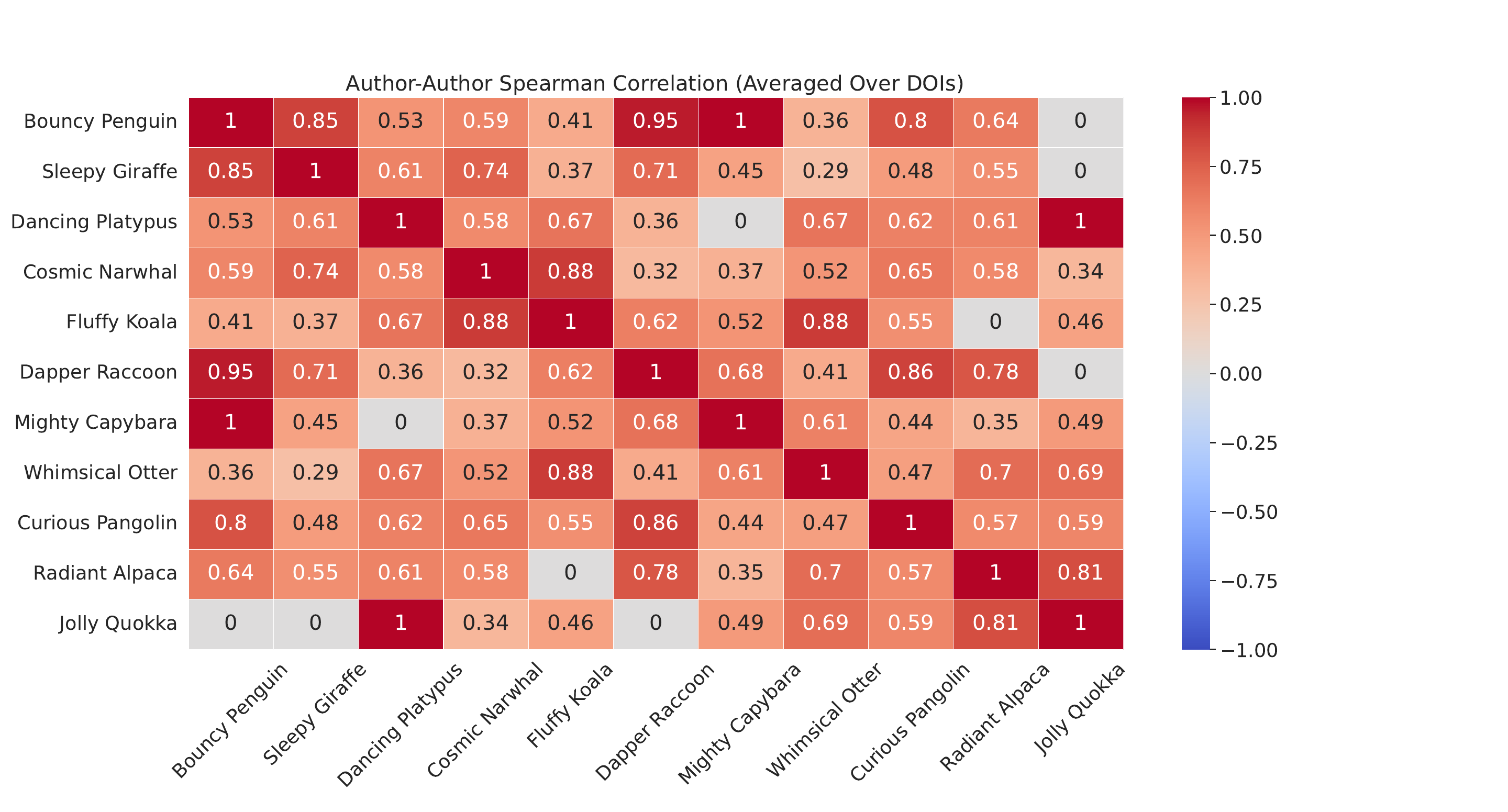}
    \caption{\textbf{Inter-annotator agreement among human labelers.} Each row and column represents a human annotator. The color scale indicates the Spearman rank correlation coefficient computed over reported power conversion efficiencies. While the diagonal (self-correlation) is perfect by definition, off-diagonal agreement is often weak, highlighting inconsistencies in manual labeling.}
    \label{fig:spearman_correlations_author_author}
\end{figure}

\subsection{Consensus dataset}

To construct a consensus baseline dataset from the manually labeled extractions, all submissions were first grouped by publication DOI, resulting in a mapping from each DOI to the list of authors and their corresponding extractions. For each DOI, the extraction containing the largest number of device entries was identified. All other extractions for the same DOI were then reordered to align with this reference extraction. To facilitate reordering, empty device entries were inserted into the annotation where necessary, and individual device entries were matched by verifying that the values of all four key parameters (FF, $V_\mathrm{OC}$, $J_\mathrm{SC}$, and PCE) matched those in the reference extraction.  

Once alignment was completed, extractions were compared across authors to identify instances in which the number of device entries matched and all four key parameters for each entry were identical. If agreement was found among at least two authors, the full extraction of one author was taken, including all additional associated fields beyond the four primary parameters. This procedure ensured that a single, consistent extraction was retained for each publication, incorporating verified information while preserving supplementary metadata where available. This resulted in a total of 15 extractions in the consensus dataset.

\subsection{Ground truth dataset}

The 15 publications from the consensus dataset were adopted to construct the ground truth dataset. All extracted entries from these publications were manually reviewed and corrected by the lead author where necessary to ensure consistency with the established annotation protocol and formatting standards. 

To improve coverage and ensure representation across a broader set of devices, compositions, and reporting formats, an additional 15 publications were manually labeled by the lead author following the same annotation protocol and formatting standards used in the initial expert labeling phase. These additions were selected to balance the dataset in terms of composition types, efficiency ranges, and underrepresented fabrication methods.

The final ground truth dataset thus comprises 30 publications, encompassing diverse perovskite formulations (e.g., methylammonium, formamidinium, cesium-based), device architectures (planar, mesoporous), and reported performance metrics (efficiencies ranging from \SI{8.1}{\percent} to \SI{23.7}{\percent}). 20 publications were used for testing, and 10 were reserved for development.

\section{LLM-based structured data extraction} \label{sec:extraction_pipeline}

\subsection{Extraction process}
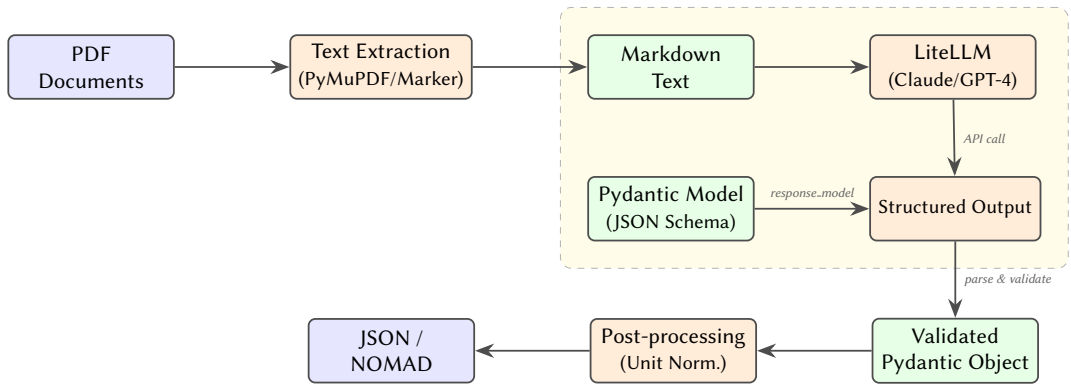
\begin{figure}[!htb]
    \centering
    \resizebox{\textwidth}{!}{
\begin{tikzpicture}[
    node distance=1.4cm and 1.8cm,
    box/.style={rectangle, draw=black!70, fill=blue!10, thick, 
                minimum width=2.6cm, minimum height=1cm, align=center, 
                rounded corners=3pt, font=\small},
    process/.style={rectangle, draw=black!70, fill=orange!15, thick,
                    minimum width=2.6cm, minimum height=1cm, align=center,
                    rounded corners=3pt, font=\small},
    data/.style={rectangle, draw=black!70, fill=green!10, thick,
                 minimum width=2.6cm, minimum height=1cm, align=center,
                 rounded corners=3pt, font=\small},
    arrow/.style={-{Stealth[length=3mm]}, thick, draw=black!70},
    label/.style={font=\footnotesize\itshape, text=black!60}
]

\node[box] (pdf) {PDF\\Documents};
\node[process, right=of pdf] (preprocess) {Text Extraction\\{\footnotesize(PyMuPDF/Marker)}};
\node[data, right=of preprocess] (text) {Markdown\\Text};
\node[process, right=of text] (litellm) {LiteLLM\\{\footnotesize(Claude/GPT-4)}};

\node[data, below=1.2cm of text] (schema) {Pydantic Model\\{\footnotesize(JSON Schema)}};
\node[process, right=of schema] (instructor) {\footnotesize{Structured Output}};
\node[data, below=1.2cm of instructor] (validated) {Validated\\Pydantic Object};

\node[process, left=of validated] (postprocess) {Post-processing\\{\footnotesize(Unit Norm.)}};
\node[box, left=of postprocess] (output) {JSON /\\NOMAD};

\draw[arrow] (pdf) -- (preprocess);
\draw[arrow] (preprocess) -- (text);
\draw[arrow] (text) -- (litellm);

\draw[arrow] (litellm) -- node[right, label] {\tiny{API call}} (instructor);
\draw[arrow] (schema) -- node[above, label] {\tiny{response\_model}} (instructor);
\draw[arrow] (instructor) -- node[right, label] {\tiny{parse \& validate}} (validated);

\draw[arrow] (validated) -- (postprocess);
\draw[arrow] (postprocess) -- (output);

\begin{scope}[on background layer]
    \node[fit=(litellm)(instructor)(schema), fill=yellow!10, 
          draw=black!30, dashed, rounded corners=5pt, 
          inner sep=12pt, label={[font=\footnotesize]above right:instructor.from\_litellm()}] {};
\end{scope}

\end{tikzpicture}}

    \caption{\textbf{Overview of the LLM-based extraction process.} Our codebase implements a modular extraction process. Text from PDFs is extracted and then sent to an LLM with a structured output call. Some automatic validation might lead to retries. After that, a Pydantic data model is post-processed for an upload to NOMAD.}
    \label{fig:extraction-flow}
\end{figure}

A sketch of our extraction process is shown in \Cref{fig:extraction-flow}.
For all extractions, we use inference temperature 0 (greedy decoding). In case the extracted output cannot be converted into a valid Pydantic model, we perform at maximum three extraction retries with the \texttt{instructor} library, where the error (history) is appended to the request.

\subsection{Automated validation}
LLMs can occasionally invent plausible-looking numbers---a phenomenon known as hallucination---or misinterpret complex articles. In addition there might also be problems in the extraction that cause inconsistencies. Our pipeline addresses that.

\subsubsection{The \enquote{hallucination} search}
Every time the model extracts a value (e.g., an efficiency of \SI{21.3}{\percent}), the system searches the original PDF to verify that this number actually exists in the text. Because authors report numbers in various formats, the search is flexible. It accepts the value if it appears as:
\begin{itemize}
    \item \textbf{An exact match:} Finding  \enquote{21.3} or \enquote{21.30}.
    \item \textbf{A rounded integer:} Finding \enquote{21\%} for an extracted \SI{21.3}{\percent}.
    \item \textbf{A fraction:} Finding \enquote{0.213} for an extracted \SI{21.3}{\percent}.
    \item \textbf{A scaled unit:} Finding \enquote{548} (millivolts) for an extracted 0.548 (volts).
\end{itemize}
If the system cannot locate the number in any of these forms, it flags the value as a hallucination and removes it.

\subsubsection{Physical consistency checks}
\label{sec:physics_consistency}

\begin{figure}[!htb]
    \centering    \includegraphics[width=\linewidth]{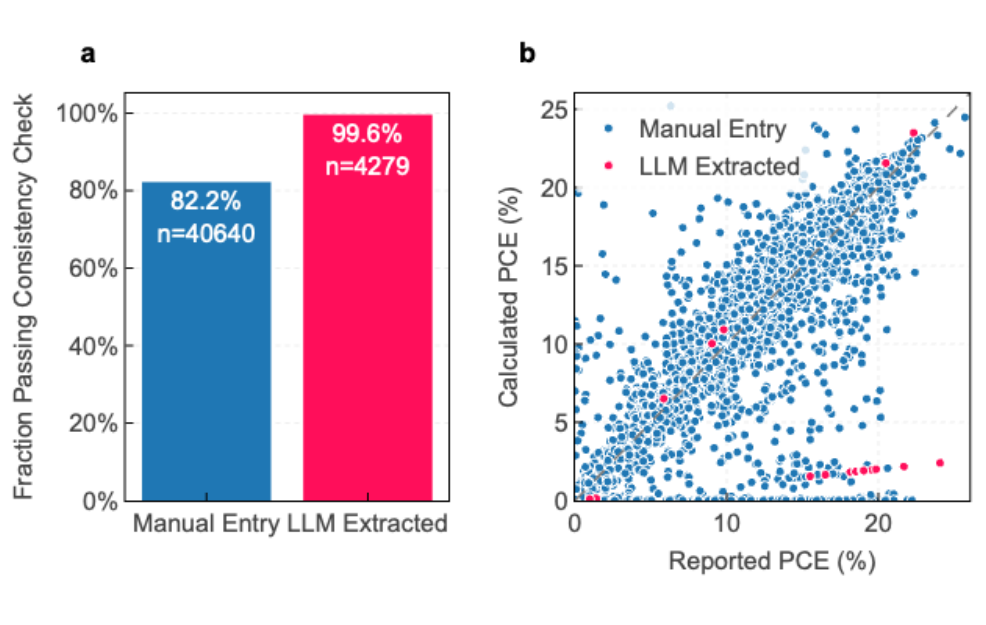}
    \caption{\textbf{Physics-based consistency validation of photovoltaic performance metrics.}
    \textbf{a.} Fraction of solar cell entries satisfying the PCE consistency relation
    $\mathrm{PCE} \approx J_\mathrm{sc} \times V_\mathrm{oc} \times \mathrm{FF} / P_\mathrm{in}$
    within an absolute tolerance of \SI{0.2}{\percent}.
    Manual literature entries pass this consistency check in 82.2\% of cases ($n=40{,}640$), reflecting heterogeneous reporting practices and aggregation artifacts in legacy curation.
    In contrast, 99.6\% of LLM-extracted entries ($n=4{,}279$) satisfy the constraint, as the consistency check is enforced during automated ingestion.
    \textbf{b.} Calculated versus reported power conversion efficiency for individual devices that \textbf{do not} pass the consistency check.
    Manual entries exhibit substantial scatter and frequent large deviations from the identity line, indicating internally inconsistent combinations of $J_\mathrm{sc}$, $V_\mathrm{oc}$, FF, and PCE.
    Outliers in the LLM-extracted data predominantly correspond to cases with  incorrectly resolved illumination metadata, leading to systematic underestimation of calculated efficiencies.
   This analysis was restricted to devices with explicitly reported illumination intensities close to standard 1~sun conditions.}
    \label{fig:physics_consistency_check}
\end{figure}

We also enforce a physics-based internal consistency check on extracted photovoltaic performance metrics.
For single-junction solar  the power conversion efficiency (PCE) is defined as
\begin{equation}
    \mathrm{PCE} = \frac{J_\mathrm{sc} \times V_\mathrm{oc} \times \mathrm{FF}}{P_\mathrm{in}},
\end{equation}
where $J_\mathrm{sc}$ is the short-circuit current density, $V_\mathrm{oc}$ the open-circuit voltage, FF the fill factor, and $P_\mathrm{in}$ the incident power density.
Using the units adopted throughout this work ($J_\mathrm{sc}$ in \unit{mA\,cm^{-2}}, $V_\mathrm{oc}$ in \unit{V}, FF in percent) and assuming standard AM1.5G illumination ($P_\mathrm{in}=\SI{100}{mW\,cm^{-2}}$), this reduces to
\begin{equation}
    \mathrm{PCE}_\mathrm{calc}(\%) = \frac{J_\mathrm{sc} \times V_\mathrm{oc} \times \mathrm{FF}}{100}.
\end{equation}

For each extracted device where all four quantities are available, we compute $\mathrm{PCE}_\mathrm{calc}$ and compare it to the reported PCE.
Entries are discarded if the absolute deviation exceeds \SI{0.2}{\percent}.
This tolerance accommodates rounding and reporting precision while effectively eliminating transcription errors, spurious numerical combinations, and occasional hallucinated values. We note that a significant fraction of the devices curated in the legacy perovskite database do not fulfill this consistency check (see \Cref{fig:physics_consistency_check}). This might also be related to the fact that on many occasions, these values were curated as \enquote{average} values from a given set of devices. 

This check is particularly important for the extraction pipeline in publications reporting multiple devices.
In such cases, performance metrics are often distributed across several paragraphs, tables, or figures, and both manual and automated curation can inadvertently combine values originating from different devices.
Enforcing the $J_\mathrm{sc}$--$V_\mathrm{oc}$--FF--PCE relation therefore acts as a device-level coherence constraint, ensuring that extracted quantities correspond to a single, internally consistent measurement context.

Beyond resolving ambiguities in multi-device reports, this filter also removes a substantial number of inconsistencies present in legacy manually curated data.
These inconsistencies frequently arise from heterogeneous reporting practices, transcription errors in the manual curation, or the use of $J_\mathrm{sc}$ values derived from integrated EQE measurements rather than from the same current--voltage scan used to determine $V_\mathrm{oc}$ and FF.

A small fraction of LLM-extracted entries also fail this check. 
In nearly all such cases, the deviation can be traced back to incorrectly resolved illumination metadata during extraction.
For example, assuming a standard incident power density for measurements performed under non-standard illumination intensities leads to systematically underestimated calculated efficiencies. Work is in progress to include the extracted illumination conditions in the filter check, making it more robust.

Finally, we exclude tandem solar cells and photovoltaic modules, which operate in distinct physical regimes, by applying empirical upper bounds of
$V_\mathrm{oc} < \SI{1.56}{\volt}$ and $\mathrm{PCE} < \SI{27.5}{\percent}$.
This additional filtering step maximizes the likelihood that the resulting dataset contains only single-junction perovskite solar cells and remains physically homogeneous.

\subsection{Evaluation pipeline}

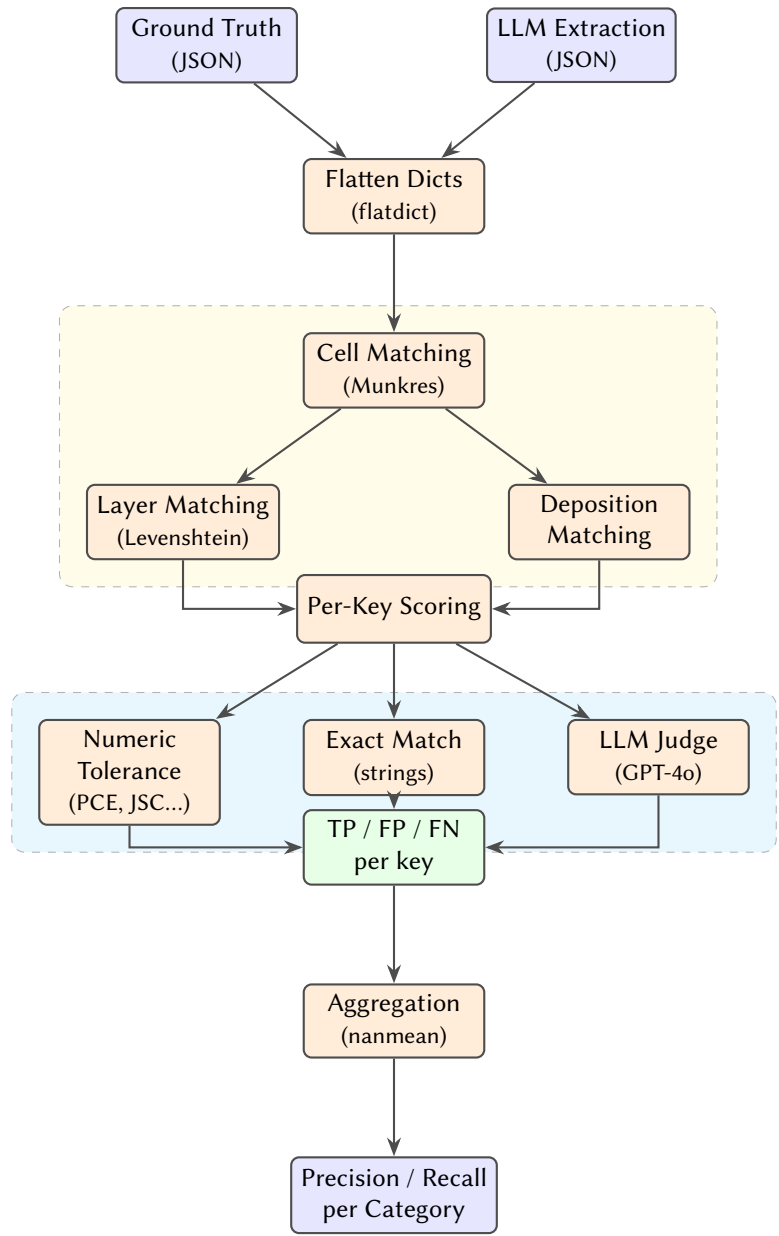
\begin{figure}[!htb]
    \centering
    \begin{tikzpicture}[
    node distance=1.3cm and 1.6cm,
    box/.style={rectangle, draw=black!70, fill=blue!10, thick, 
                minimum width=2.4cm, minimum height=0.9cm, align=center, 
                rounded corners=3pt, font=\small},
    process/.style={rectangle, draw=black!70, fill=orange!15, thick,
                    minimum width=2.4cm, minimum height=0.9cm, align=center,
                    rounded corners=3pt, font=\small},
    data/.style={rectangle, draw=black!70, fill=green!10, thick,
                 minimum width=2.4cm, minimum height=0.9cm, align=center,
                 rounded corners=3pt, font=\small},
    arrow/.style={-{Stealth[length=2.5mm]}, thick, draw=black!70},
    label/.style={font=\scriptsize\itshape, text=black!60}
]

\node[box] (truth) {Ground Truth\\{\footnotesize(JSON)}};
\node[box, right=2.5cm of truth] (extraction) {LLM Extraction\\{\footnotesize(JSON)}};

\node[process, below=1.5cm of $(truth)!0.5!(extraction)$] (flatten) {Flatten Dicts\\{\footnotesize(flatdict)}};
\node[process, below=of flatten] (cellmatch) {Cell Matching\\{\footnotesize(Munkres)}};
\node[process, below left=1cm and 0.3cm of cellmatch] (layermatch) {Layer Matching\\{\footnotesize(Levenshtein)}};
\node[process, below right=1cm and 0.3cm of cellmatch] (deposmatch) {Deposition\\Matching};

\node[process, below=2.2cm of cellmatch] (scoring) {Per-Key Scoring};
\node[process, below left=1cm and 1cm of scoring] (tolerance) {Numeric\\Tolerance\\{\footnotesize(PCE, JSC...)}};
\node[process, below=1cm of scoring] (exact) {Exact Match\\{\footnotesize(strings)}};
\node[process, below right=1cm and 1cm of scoring] (llmjudge) {LLM Judge\\{\footnotesize(GPT-4o)}};

\node[data, below=2.2cm of scoring] (metrics) {TP / FP / FN\\per key};
\node[process, below=of metrics] (aggregate) {Aggregation\\{\footnotesize(nanmean)}};
\node[box, below=of aggregate] (output) {Precision / Recall\\per Category};

\draw[arrow] (truth) -- (flatten);
\draw[arrow] (extraction) -- (flatten);

\draw[arrow] (flatten) -- (cellmatch);
\draw[arrow] (cellmatch) -- (layermatch);
\draw[arrow] (cellmatch) -- (deposmatch);
\draw[arrow] (layermatch) |- (scoring);
\draw[arrow] (deposmatch) |- (scoring);

\draw[arrow] (scoring) -- (tolerance);
\draw[arrow] (scoring) -- (exact);
\draw[arrow] (scoring) -- (llmjudge);
\draw[arrow] (tolerance) |- (metrics);
\draw[arrow] (exact) -- (metrics);
\draw[arrow] (llmjudge) |- (metrics);

\draw[arrow] (metrics) -- (aggregate);
\draw[arrow] (aggregate) -- (output);

\begin{scope}[on background layer]
    \node[fit=(cellmatch)(layermatch)(deposmatch), fill=yellow!10, 
          draw=black!30, dashed, rounded corners=5pt, 
          inner sep=10pt, label={[font=\scriptsize]above:Optimal Assignment}] {};
    \node[fit=(tolerance)(exact)(llmjudge), fill=cyan!8, 
          draw=black!30, dashed, rounded corners=5pt, 
          inner sep=10pt, label={[font=\scriptsize]below:DeepDiff + LLM}] {};
\end{scope}

\end{tikzpicture}

    \caption{\textbf{Evaluation pipeline for LLM-based extraction.} Ground truth and extracted JSON structures are flattened and aligned using the Munkres algorithm for optimal bipartite matching at three hierarchical levels: solar cells, layers, and deposition steps. Layer matching uses Levenshtein distance on functionality and name fields. Per-key scoring applies tolerance-based comparison for numeric fields (PCE, JSC, VOC, FF) and exact matching for strings; mismatches on string fields are re-evaluated using an LLM-as-judge (GPT-4o) for semantic equivalence. True positives (TP), false positives (FP), and false negatives (FN) are aggregated per key and averaged across categories (composition, deposition, stability, etc.) to compute precision and recall metrics.}
    \label{fig:validation_flow}
\end{figure}

To quantitatively assess the performance of our LLM-based extraction pipeline, we developed an evaluation framework that compares the structured JSON output against manually curated ground truth data (see \Cref{fig:validation_flow}). 
The framework is designed to robustly handle the nested complexity of perovskite solar cell data and potential discrepancies in the ordering of extracted entities. 
We normalize both datasets using \texttt{pint} for unit conversion, then match each extracted solar cell with exactly one corresponding ground truth entity using hierarchical optimal matching (yellow in \Cref{fig:validation_flow}).
After matching devices, we compute metrics per field.
We evaluate matched extractions using precision and recall metrics to quantify overall performance. Additionally, we calculate precision scores for each individual key in the extraction data. When a key is determined to be incorrect, we employ LLM-as-a-judge as a fallback mechanism to provide secondary validation and assessment.

\subsubsection{Hierarchical optimal matching}
The multi-level matching process  uses the Munkres algorithm,\autocite{munkres1957algorithms, kuhn1955hungarian} a solution to the assignment problem, to find the optimal pairing between entities in the ground truth and the LLM extraction for a fair evaluation. 
This matching occurs sequentially at the device, layer, and deposition levels.

\paragraph{Device-level matching} First, we match solar cell devices between the ground truth ($D_{\text{truth}}$) and the extraction ($D_{\text{ext}}$). For each potential pair of a truth device $i$ and an extracted device $j$, we compute a dissimilarity score, $C_{ij}$, which forms a cost matrix. 
The Munkres algorithm then minimizes the total cost of assignments. The score $C_{ij}$ is a weighted sum of three components:
$$
C_{ij} = w_1 S_{\text{func}}(i, j) + w_2 S_{\text{struct}}(i, j) + w_3 S_{\text{depo}}(i, j)
$$
Here, $S_{\text{func}}$ is a dissimilarity score ($w_1=0.7$) based on the Levenshtein ratio of the material names within each functional layer (e.g., perovskite, electron transport layer). $S_{\text{struct}}$ ($w_2=0.1$) and $S_{\text{depo}}$ ($w_3=0.2$) are dissimilarity scores derived from the \texttt{DeepDiff} library,\autocite{Dehpour_DeepDiff} quantifying differences in the overall device JSON structure and the sequence of deposition methods, respectively.

\paragraph{Layer and deposition matching} After establishing optimal device pairs, a similar matching process is recursively applied within each pair. Constituent layers are matched based on the Levenshtein distance between concatenated strings of their functionality and material name. Subsequently, within each matched layer, the individual deposition steps are aligned based on the Levenshtein distance of their method and temperature parameters. This hierarchical approach ensures that comparisons are always made between the most plausible corresponding entities. Any ground truth devices that remain unmatched after this process are counted towards device-level false negatives, contributing to a lower device recall.

\subsubsection{Precision and recall calculation}

Following the matching process, we calculate the precision and recall for each field. 
A comprehensive breakdown of the scoring logic, tolerances, and semantic matching criteria for every data field is detailed below. To facilitate comparison, the nested JSON structure of each matched device pair is flattened into a key-value format (e.g., \texttt{layers:0:name}). 
Prior to evaluation, all numerical data from both the ground truth and the extraction were canonicalized to ensure consistent units and formats.

\paragraph{Unit normalization} 
To ensure consistent and comparable numerical values during evaluation, we applied a unit normalization procedure to both the ground truth and extracted data before scoring. 
This process recursively traverses the nested JSON structure, identifying fields that contain both a numerical \texttt{value} and an associated \texttt{unit}. 
Using the \texttt{pint} library\autocite{pint}, each quantity is parsed, and its physical dimensionality is determined (e.g., voltage, current density, temperature). 
A predefined mapping from dimensionality to canonical units is then used to convert values into standard forms. For example, current densities are converted to \unit{mA/cm^2}, voltages to \unit{V}, efficiencies and fill factors to percentages (\%), and temperatures to degrees Celsius (\unit{\degreeCelsius}). More specialized quantities such as irradiance (\unit{mW/cm^2}), concentration (\unit{mol/L}), and thickness (\unit{nm}) are also handled. Time values (e.g., \enquote{2 days}) are internally converted to seconds for consistency. This normalization ensures that differences in unit expression do not affect evaluation outcomes. 

\clearpage

\subsection{Metrics}

\paragraph{Recall} measures the fraction of fields present in the ground truth that were successfully identified by the pipeline. 
A field is considered found if its key exists in the extraction and its value is not null. The per-device recall is the number of found fields divided by the total number of fields in the ground truth device. Devices that were not matched to any extraction receive a recall score of 0.

\paragraph{Precision} measures the accuracy of the extracted values. For every field present in both the ground truth and the matched extraction, we assess if the extracted value is correct based on the field type.
\begin{itemize}
    \item \textbf{Numerical Fields:} Key performance metrics are evaluated against absolute tolerances: Power Conversion Efficiency ($\pm$\SI{0.1}{\percent}), Short-Circuit Current Density ($\pm$\SI{0.1}{mA/cm^2}), Open-Circuit Voltage ($\pm$\SI{0.01}{V}), and Fill Factor ($\pm$\SI{0.1}{\percent}). All other numerical values are assessed with a relative tolerance of \SI{1}{\percent}.
    \item \textbf{String-Based Fields:} For textual data, correctness is determined using a multi-step process involving direct matching and, where necessary, semantic evaluation by an LLM judge.
\end{itemize}

\paragraph{LLM-as-Judge} 
For qualitative fields where simple string comparison is insufficient, we employed an LLM-as-Judge framework \autocite{li2024llmsasjudges, rios-garcia2025llmasjudge}. This is crucial for handling valid representational variants in scientific text, such as chemical formulas (e.g., \enquote{\ce{MAPbI3}} vs.\ \enquote{\ce{CH3NH3PbI3}}), material abbreviations with typos (e.g., \enquote{Spiro-MeOTAD} vs. \enquote{Spiro-OMeTAD}), or descriptive parameters (e.g., \enquote{10 min \ce{TiCl4} treatment} vs. \enquote{\ce{TiCl4} treatment for 10 min}). 
To ensure the reliability of this approach, we benchmarked several candidate models on a manually curated set of 43 challenging comparisons. 
GPT-4o was selected as the judge for all final evaluations due to its superior performance, achieving 97\% accuracy on this benchmark, significantly outperforming other models tested.

In implementation, we used the \texttt{Instructor} Python library\autocite{Liu_Instructor} to issue calls to GPT-4o, providing the model with the ground truth value, the extracted value, and a natural language prompt instructing it to return a binary judgment---\texttt{True} if the values were semantically equivalent, and \texttt{False} otherwise. The prompt also included context about the field and device to assist the model in making accurate determinations. 

\begin{mintedbox}{python}
class Judgement(BaseModel):
        judgement: bool = Field(
            None,
            description="The final say whether the given values match (TRUE) or not (FALSE).",
        )
    prompt_messages = [
        {
            "role": "system",
            "content": "You are an expert scientist that judges whether two provided values match in a data extraction evaluation routine for perovskite solar cells. You will be given the whole ground truth, the value in the ground truth, and the value from the extraction. You have to check if the two values match conceptually. They do not have to be exactly the same. Formulas can be variable. c-TiO2 is the same as TiO2. Only respond with either TRUE or FALSE",
        },
        {
            "role": "user",
            "content": f"Complete ground truth: {str(ground_truth)}\n Truth value: {str(value_truth)} \n Extraction value: {str(value_extraction)}",
        },
    ]
\end{mintedbox}

\paragraph{Micro-averaged evaluation metrics} 
The final per-field precision and recall are calculated by aggregating the counts of True Positives (TP), False Positives (FP), and False Negatives (FN) across all documents in the test set, according to the standard definitions:
\begin{equation*}
\text{Precision} = \frac{\text{TP}}{\text{TP} + \text{FP}}
\quad ; \quad
\text{Recall} = \frac{\text{TP}}{\text{TP} + \text{FN}}
\end{equation*}

\paragraph{Model comparisons}

To evaluate model performance across all extracted fields, we report micro-averaged precision and recall. In this setting, true positives (TP), false positives (FP), and false negatives (FN) are aggregated across all field types and all files prior to score computation, so every individual prediction carries equal weight regardless of field identity (e.g., efficiency, voltage, material names, or deposition parameters). These scores therefore reflect the raw performance of the underlying model alone. In practice, overall pipeline performance is expected to be higher, as downstream filtering and consistency checks mitigate spurious predictions. The development set comprises 35 devices derived from 10 extractions, while the test set consists of 101 devices from 20 extractions.

We used the following model checkpoints: 
\begin{itemize}
    \item \texttt{claude-opus-4-1-20250805}
    \item \texttt{claude-opus-4-20250514}
    \item \texttt{claude-sonnet-4-20250514}
    \item \texttt{gpt-4.1-2025-04-14}
    \item \texttt{gpt-4o-2024-08-06}
    \item \texttt{gpt-5-2025-08-07}
    \item \texttt{gpt-5-mini-2025-08-07}
\end{itemize}

\begin{figure}[htb]
    \centering
    \includegraphics[width=.7\linewidth]{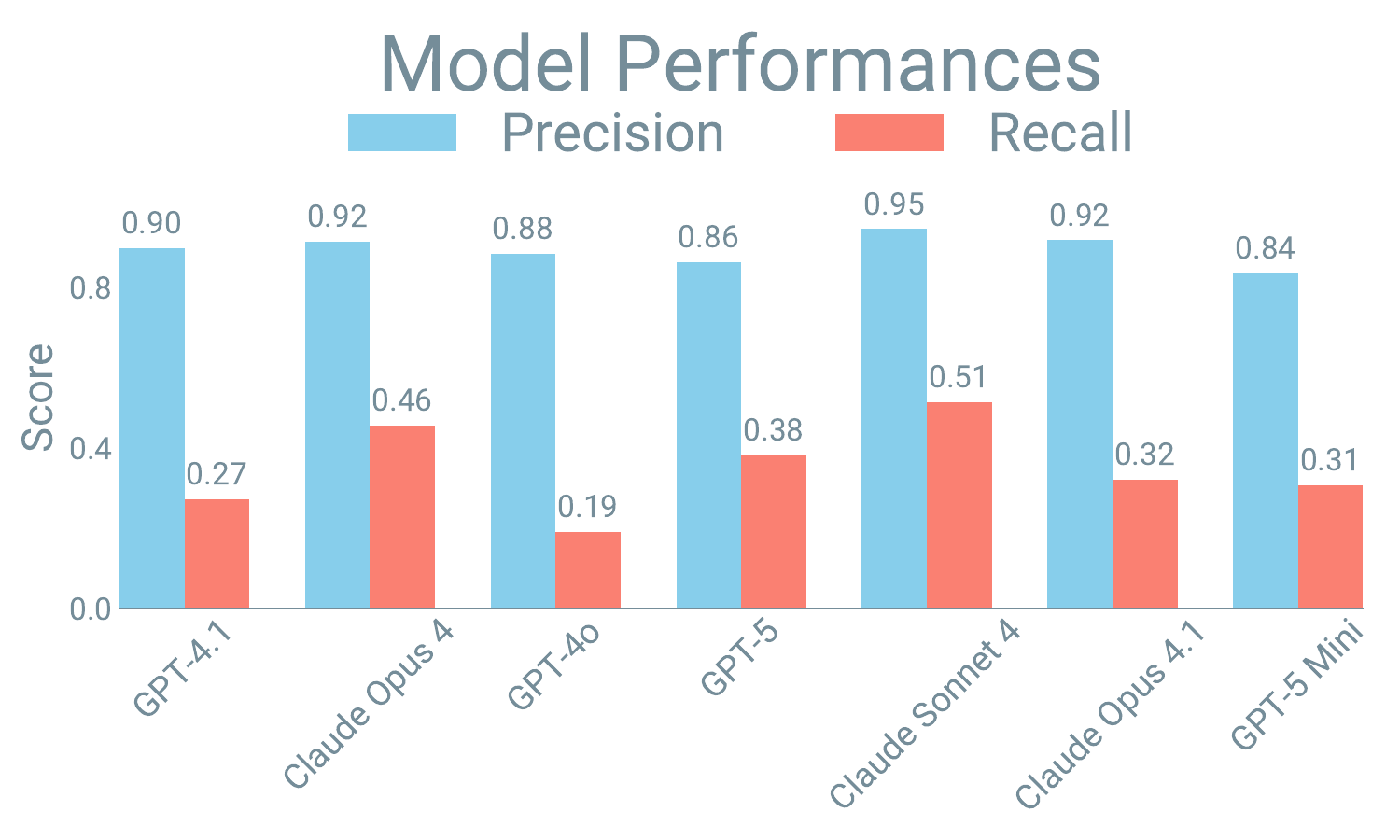}
    \caption{\textbf{Comparison of the performance of different models on the development set.} \textbf{a} Precision. \textbf{b} Recall.}
    \label{fig:model_comparison}
\end{figure}

\begin{figure}
    \centering
    \includegraphics[]{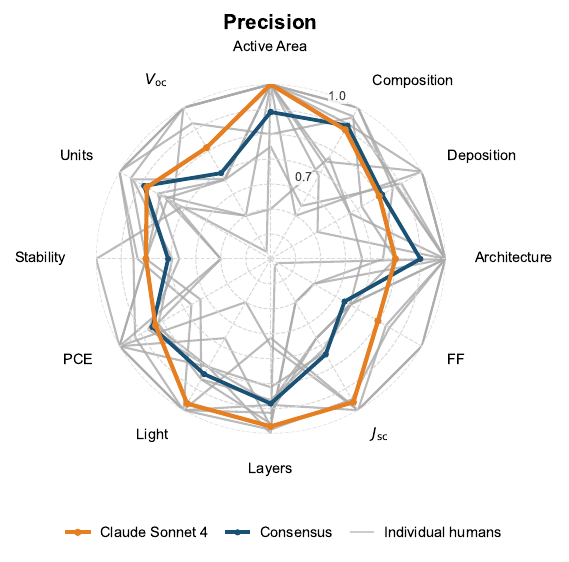}
    \caption{\textbf{Benchmarking autonomous extraction against human expert performance.} Radar plot comparing the precision of individual human domain experts (grey lines), an expert consensus (dark blue line), and the LLM-based pipeline (orange line) across 13 extraction categories. While individual human annotators exhibit significant variance, the LLM matches or exceeds the precision of the top-tier human experts. The proximity of the LLM trace to the expert consensus across both numerical (e.g., $V_{oc}$, PCE) and categorical (e.g., architecture, layers) parameters demonstrates the system's reliability as a high-fidelity surrogate for manual curation.}
    \label{fig:comparison_with_human_performance}
\end{figure}

\input{detailed_tables/score_tables}

\clearpage
\section{N-I-P material distribution}

\begin{figure}[htb]
    \centering
    \includegraphics[width=1\linewidth]{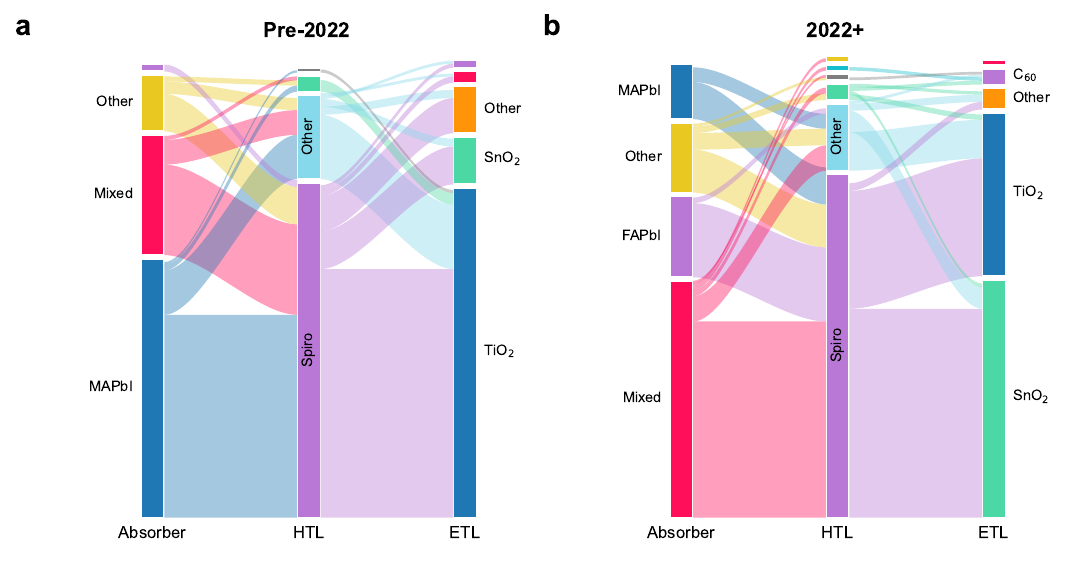}
    \caption{\textbf{Evolution of materials used in devices with n-i-p architecture.}}
    \label{fig:nip_evolution}
\end{figure}

\clearpage

\section{Distribution shift in machine learning property prediction} \label{sec:ml-case-study}
\begin{figure}[htb]
    \centering
    \includegraphics[width=\linewidth]{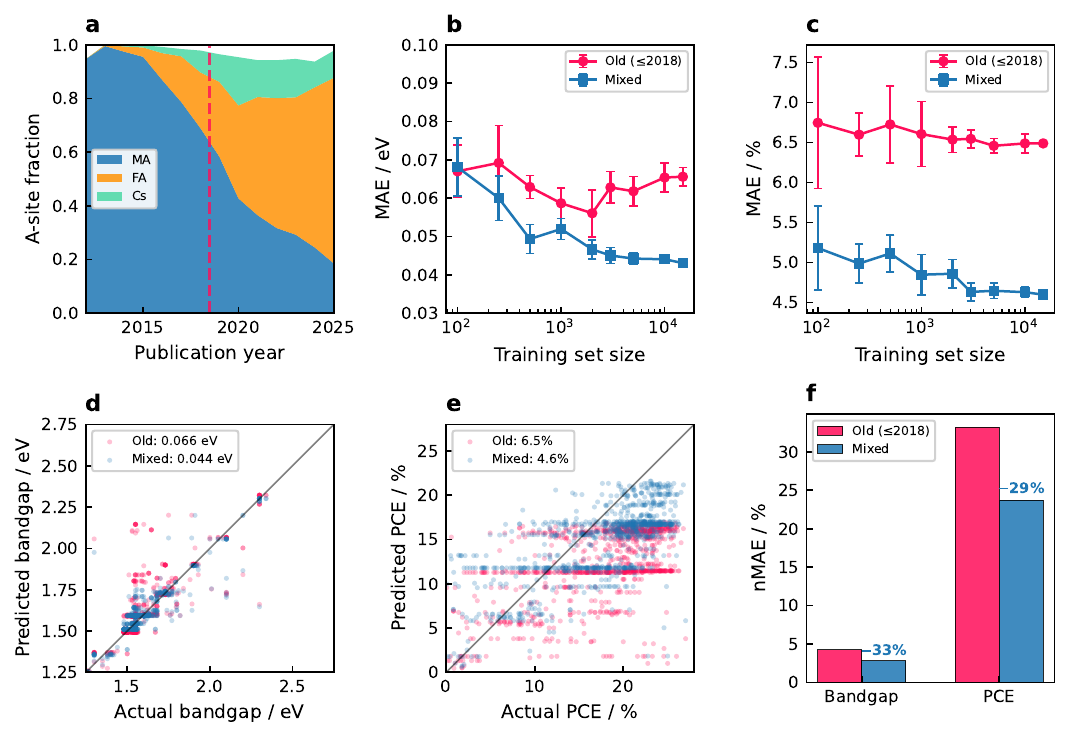}
    \caption{\textbf{Distribution shift in machine learning property prediction.}
\textbf{a.} Evolution of A-site composition showing the shift from MA-dominated (2012--2016) to FA-rich (2022--2025) perovskites. The dashed line marks the historical/modern data boundary.
\textbf{b} Learning curves for bandgap prediction comparing models trained on historical data only (Old, red) versus data spanning the full database (Mixed, blue). Error bars indicate $\pm1$ standard deviation over 5 random seeds.
\textbf{c} Learning curves for PCE prediction.
\textbf{d} Parity plot for bandgap prediction on modern test data (2022--2025).
\textbf{e} Parity plot for PCE prediction.
\textbf{f} Summary comparison of normalized mean absolute error (nMAE) for both targets. Percentages indicate relative improvement of the mixed model over the historical model.}
    \label{fig:ml-case-study}
\end{figure}

The rapid compositional evolution documented in \perla has direct implications for machine learning models trained on historical perovskite data. To quantify this effect, we performed a distribution shift analysis comparing models trained on historical data (2012–-2018) versus models trained on data spanning the full database (2012–-2025).

We extracted 11 composition-based features from the database representing normalized ion fractions at each perovskite lattice site: A-site (MA, FA, Cs, other), B-site (Pb, Sn, other), and X-site (I, Br, Cl, other). Test sets of 1,000 samples each were held out from publications dated 2022–-2025 for bandgap and power conversion efficiency (PCE) prediction. Random Forest regressors (100 trees, max depth 10) were trained on either (i) historical data only ($\le2018$, \enquote{Old}) or (ii) data spanning the full database excluding test samples (\enquote{Mixed}), with matched training set sizes of 3,000 samples. Learning curves were computed by varying training set size from 100 to 15,000 samples, averaged over 5 random seeds.

The A-site composition shifted dramatically between the historical (2012–-2018) and modern (2022-–2025) periods: methylammonium (MA) decreased from 80\% to 27\%, while formamidinium (FA) increased from 14\% to 56\% (\Cref{fig:ml-case-study}\textbf{a}). This compositional shift directly impacts model generalization. For bandgap prediction, the model trained on historical data achieved nMAE of \SI{4.2}{\percent} on modern test data, compared to nMAE of \SI{2.8}{\percent} for the model trained on mixed data---a \SI{33}{\percent} improvement. For PCE prediction, we found nMAE decreased from \SI{33.2}{\percent} to \SI{23.7}{\percent}, representing a \SI{29}{\percent} improvement (\Cref{fig:ml-case-study}\textbf{d,e,f}).

Critically, learning curve analysis revealed that this performance gap persists across all training set sizes (\Cref{fig:ml-case-study}\textbf{b,c}). Increasing the historical training set from 100 to 15,000 samples did not close the gap with the mixed model. This demonstrates that the quantity of historical data cannot compensate for a distribution mismatch.

These results underscore the importance of continuously updated databases for machine learning applications in materials science. As research frontiers evolve, training data must reflect current compositional diversity to maintain model relevance. This case study provides quantitative evidence that database maintenance is not merely an archival concern but a prerequisite for accurate property prediction in rapidly evolving materials domains.

\clearpage
\section{\perla in NOMAD} \label{sec:nomad_app}

\perla in NOMAD is built on an automated data ingestion, curation, and normalization pipeline that converts literature-reported results into fully structured and interoperable research data. This curated dataset can then be explored through an interactive graphical user interface, enabling flexible filtering and analysis across materials, device architectures, performance metrics, and publication metadata.

\subsection{From LLM-extracted data to a curated NOMAD archive}

\begin{figure}[htbp]
  \centering
  \includegraphics[width=\textwidth]{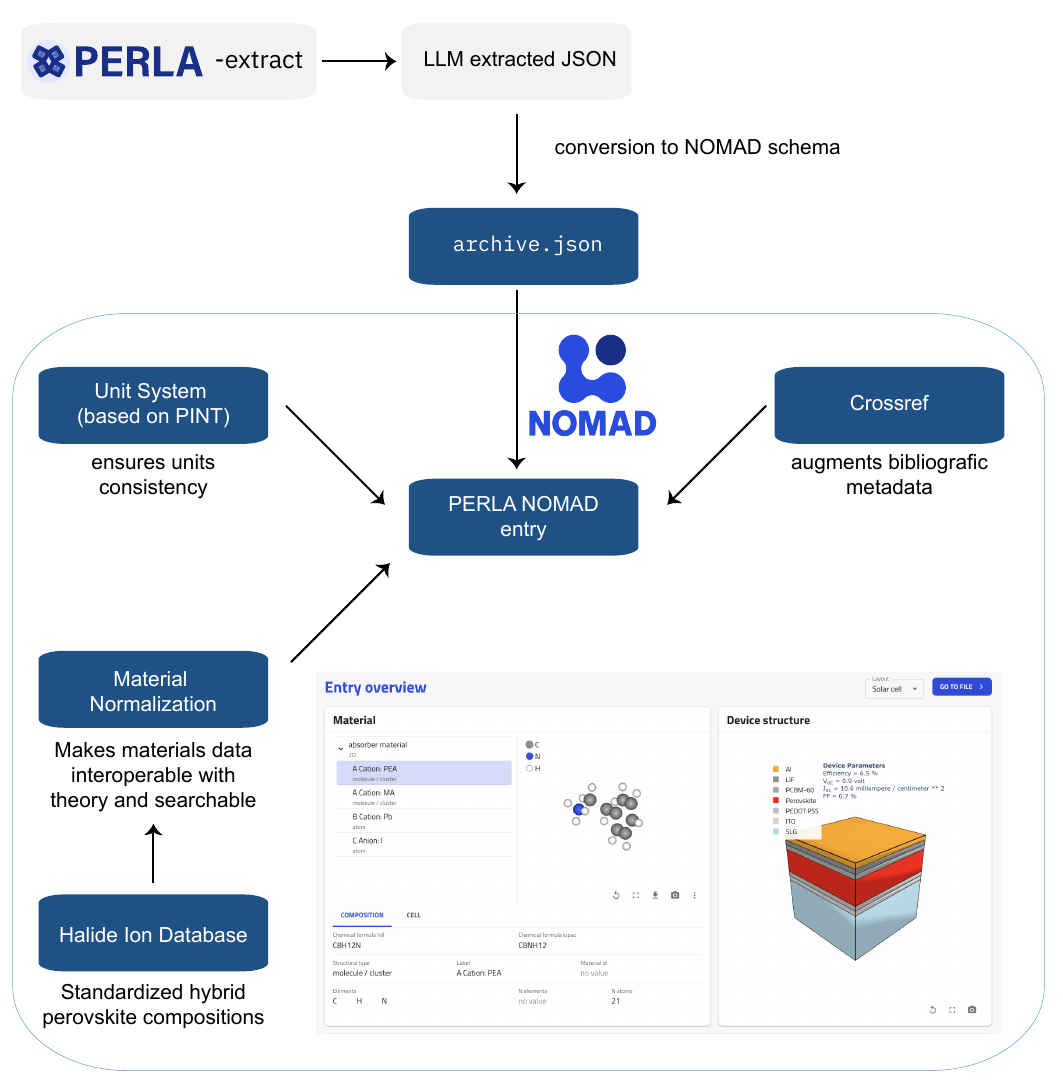}
\caption{\textbf{Workflow for integrating LLM-extracted perovskite solar cell data into NOMAD.}
Literature data extracted using large language models are converted into a NOMAD-compliant \texttt{archive.json} file and uploaded to NOMAD, triggering automated normalization and enrichment. The resulting curated dataset forms a fully integrated NOMAD entry (referred here as \perla's NOMAD entry.}
  \label{fig:perla_pipeline}
\end{figure}

In addition to the interactive exploration interface described in the following sections, the Perovskite Solar Cell Database in NOMAD is underpinned by an automated data ingestion and normalization pipeline that converts literature-extracted information into fully curated NOMAD entries. This process is illustrated schematically in \Cref{fig:perla_pipeline}.

Structured data is first extracted from the scientific literature using large language models (LLMs) and stored in intermediate JSON representations. These JSON files are subsequently converted into a NOMAD-compliant \texttt{archive.json} format that adheres to the NOMAD schema definition for the LLM extracted data entries. This conversion step ensures that the extracted information is expressed in a machine-readable and semantically well-defined form before ingestion.

Once uploaded to NOMAD, the archive automatically triggers a sequence of normalization and enrichment procedures. Hybrid perovskite compositions are standardized using the Halide Perovskite Ions Database, \autocite{maqsood_towards_2025} enabling consistent representation of mixed-cation and mixed-halide systems. Bibliographic metadata are augmented through automated queries to the Crossref API, while physical quantities are harmonized using NOMAD’s unit system based on the \texttt{pint} library, ensuring unit consistency across the dataset. Additional materials normalization steps extract elemental information and align materials descriptors with NOMAD’s metainfo,\autocite{ghiringhelli_shared_2022} making the data interoperable with theoretical datasets and fully searchable within the platform.

The outcome of this workflow is a curated NOMAD entry that integrates materials information, device architecture, performance metrics, stability data, and bibliographic metadata. Additional entry mapping to the original Perovskite Database in NOMAD is created, making the LLM-extracted data and the former manually curated one fully interoperable. Each perovskite solar cell entry can be explored through the NOMAD graphical user interface, linked to other NOMAD services, and reused for large-scale data analysis and machine learning applications.

\subsection{Interactive exploration of the Perovskite Solar Cell Database}

The main search and exploration interface for the Perovskite Solar Cell Database within NOMAD is available at
\url{https://nomad-lab.eu/prod/v1/staging/gui/search/perovskite-solar-cells-database/}.
The interface is designed to enable interactive and flexible exploration of experimental perovskite solar cell data curated through the PERLA pipeline.

The left-hand filter panel provides a structured set of filters that allow users to refine the dataset according to multiple criteria. These include, among others, the perovskite absorber composition, device architecture, fabrication details, publication metadata, and reported stability information. This filtering concept enables targeted queries across a highly heterogeneous experimental dataset.

The central dashboard can be configured and edited on the fly, allowing users to dynamically select different visualizations such as scatter plots, histograms, and categorical summaries. This enables rapid filtering, comparison, and high-level overview of key performance indicators and material properties without the need for external data processing.

\begin{figure}[htbp]
  \centering
  \includegraphics[width=\textwidth]{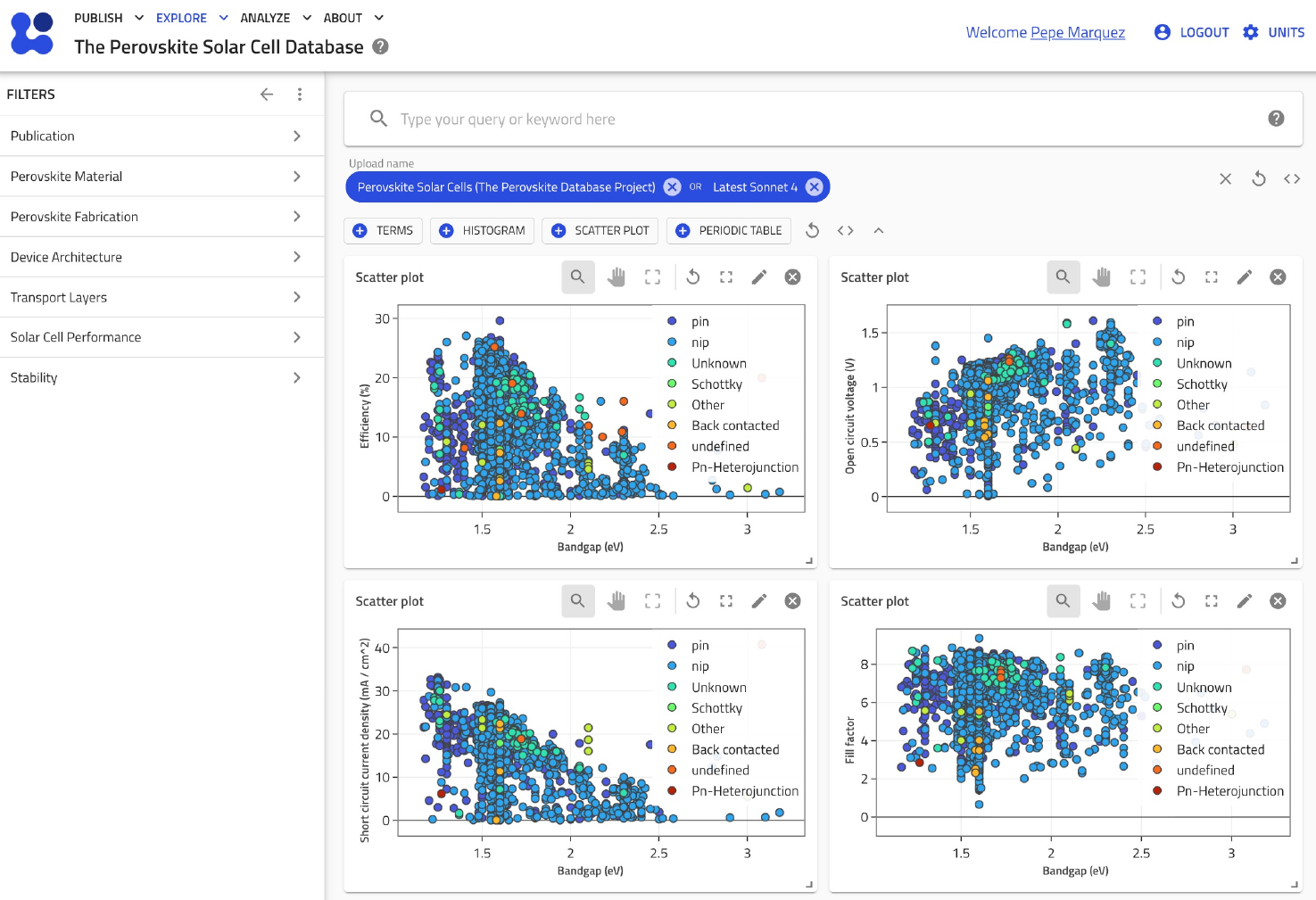}
  \caption{\textbf{Example dashboard view of the Perovskite Solar Cell Database in NOMAD.}
  The interface shows multiple scatter plots relating absorber bandgap to key device performance parameters, including efficiency, open-circuit voltage, short-circuit current density, and fill factor.}
  \label{fig:general_dashboard}
\end{figure}

\subsection{Trends in Sn-based perovskite solar cells}

\begin{figure}[htbp]
  \centering
  \includegraphics[width=\textwidth]{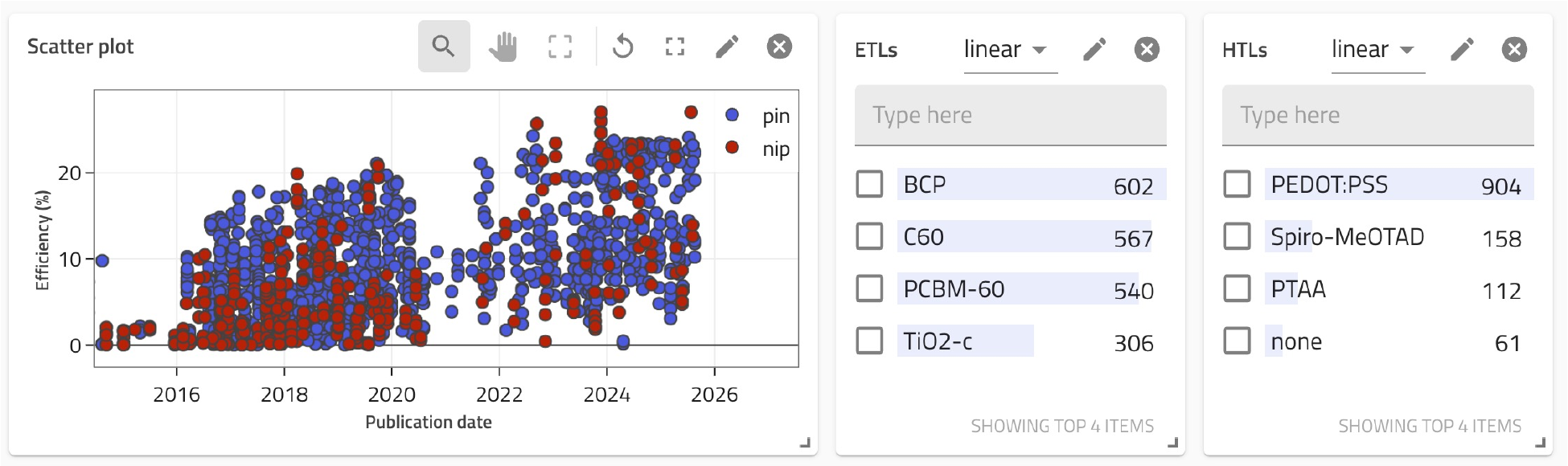}
  \caption{\textbf{Analysis of Sn-based perovskite solar cells using the NOMAD search interface.}
  Left: power conversion efficiency as a function of publication date, with data points colored by device architecture.
  Right: histograms of the four most frequently reported electron transport layers (ETLs) and hole transport layers (HTLs), highlighting dominant materials used in Sn-based device stacks.}
  \label{fig:sn_trends}
\end{figure}

The interactive NOMAD interface enables temporal and categorical analysis of specific material subclasses. As an illustrative example, \Cref{fig:sn_trends} focuses on perovskite solar cells with Sn-containing absorber layers, highlighting both performance evolution and device design choices over time.

The left panel shows the evolution of power conversion efficiency as a function of publication year, with data points colored by device architecture. A clear progressive increase in reported efficiencies can be observed over time, accompanied by an increasing prevalence of \textit{p--i--n} device architectures in more recent publications. This trend reflects the maturation of Sn-based perovskite research and the growing adoption of architectures better suited to mitigating Sn-related stability and recombination challenges.

The right panels display histograms of the most frequently used electron transport layers (ETLs) and hole transport layers (HTLs) in the filtered dataset. For ETLs, materials such as BCP, \ce{C_{60}}, PCBM-60, and compact \ce{TiO2} dominate, while PEDOT:PSS, Spiro-MeOTAD, and PTAA are the most commonly reported HTLs. These distributions provide a compact overview of prevailing material choices and facilitate rapid comparison across device stacks.

\subsection{Stability trends in FA-containing perovskites}

\begin{figure}[htbp]
  \centering
  \includegraphics[width=\textwidth]{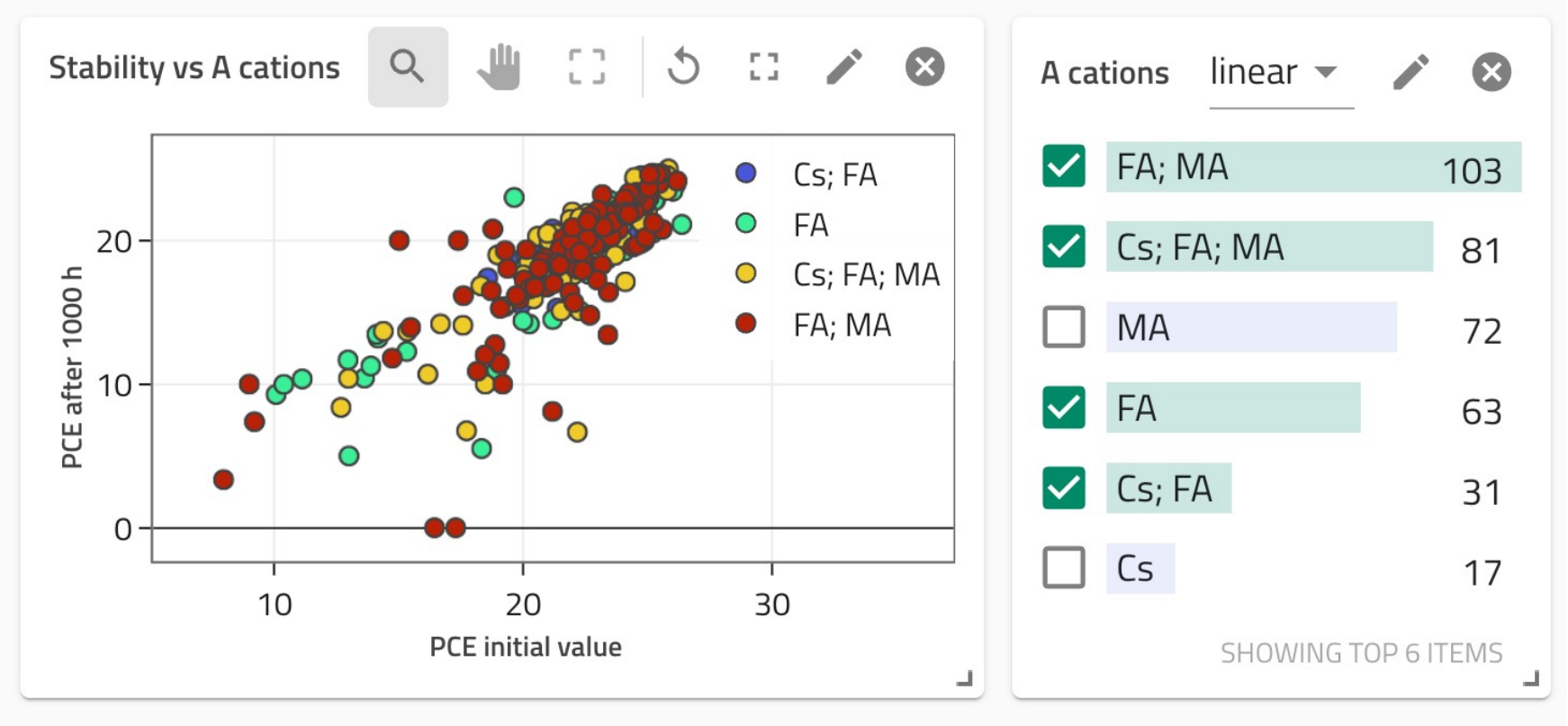}
  \caption{\textbf{Stability analysis of FA-containing perovskite solar cells using the NOMAD search interface.}
  The plot compares the initial power conversion efficiency with the efficiency measured after extended stability testing, with data points grouped by different A-site cation combinations.}
  \label{fig:fa_stability}
\end{figure}

The NOMAD database also allows stability-related data to be explored across different compositional design choices.
\Cref{fig:fa_stability} illustrates an example analysis focusing on perovskite solar cells containing formamidinium (FA) on the A-site, comparing reported initial power conversion efficiency with the efficiency retained after \SI{1000}{\hour} of extended stability testing.

The data are grouped according to different A-site cation combinations, including single-cation (FA or MA) and mixed-cation systems (FA--MA, Cs--FA, and Cs--FA--MA). When considering the dataset as a whole, no single cation combination emerges as clearly dominant in terms of stability performance. Instead, a broad scatter is observed across all groups, with substantial overlap between different compositions.

This observation highlights the multifactorial nature of stability in perovskite solar cells. Reported stability outcomes depend not only on the A-site cation chemistry but also on numerous other factors such as device architecture, transport layers, encapsulation strategies, testing protocols, and environmental conditions. As such, the analysis presented here should be understood as an intentionally simplified, high-level view of the available literature data.

\subsection{Device area and publication-level analysis}

\begin{figure}[htbp]
  \centering
  \includegraphics[width=.6\textwidth]{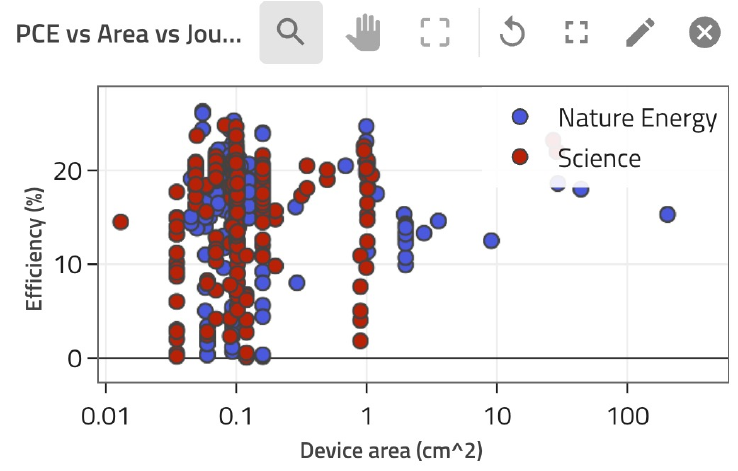}
  \caption{\textbf{Power conversion efficiency as a function of device area for perovskite solar cells published in \textit{Nature Energy} and \textit{Science}.}
  Most reported devices have active areas below \SI{1}{\centi\meter\squared}, illustrating the emphasis on small-area solar cells in the current literature.}
  \label{fig:area_publication}
\end{figure}

Beyond materials and device stack information, the database also enables analyses based on publication metadata.
\Cref{fig:area_publication} illustrates an example in which power conversion efficiency is plotted as a function of device area for perovskite solar cells reported in high-impact journals, specifically \textit{Nature Energy} and \textit{Science}.

The data show that the vast majority of reported devices in these journals have active areas below \SI{1}{\centi\meter\squared}. While high efficiencies are routinely demonstrated at these small scales, only a limited number of data points correspond to larger-area devices. This observation highlights that the primary focus of the research community, even in leading journals, remains on small-area solar cells.

At the same time, this example demonstrates how publication-related metadata in NOMAD can be leveraged to contextualize experimental results, enabling users to explore trends across journals, publication years, or authors.

\subsection{Scalable processing techniques and device area}

\begin{figure}[htbp]
  \centering
  \includegraphics[width=.8\textwidth]{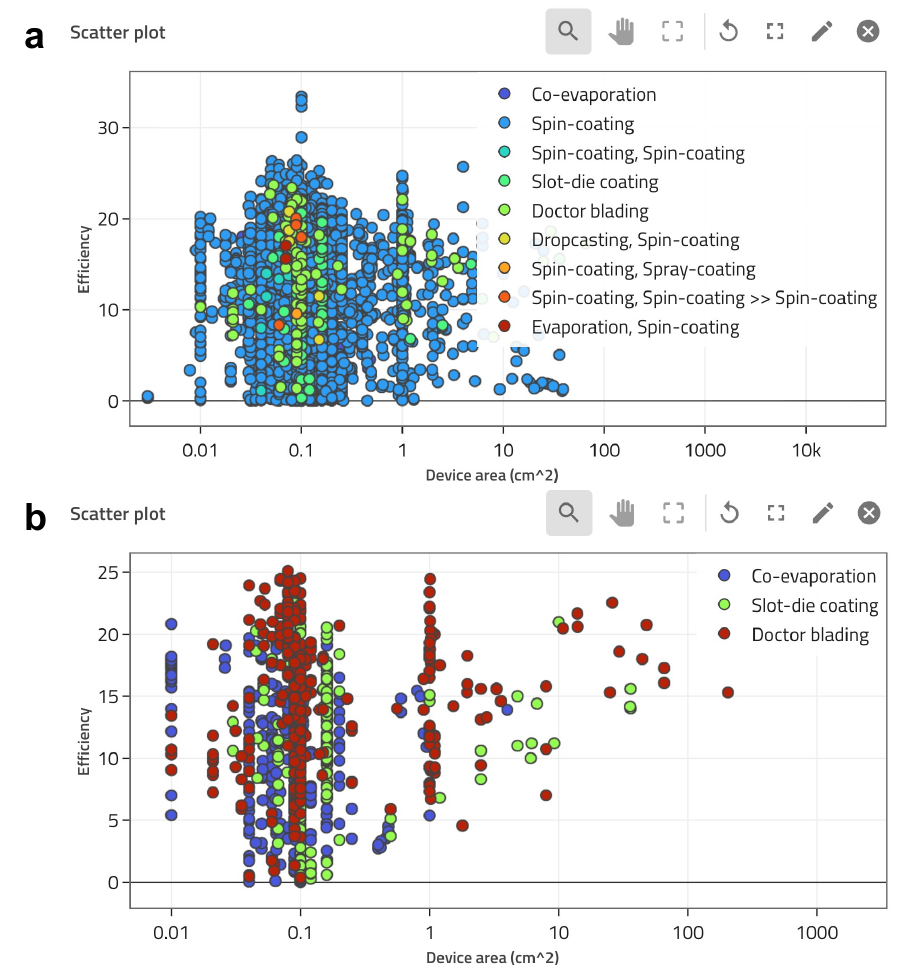}
  \caption{\textbf{Device area versus efficiency for perovskite solar cells grouped by fabrication technique.}
  \textbf{a} Overview of reported device performance across a wide range of processing routes, dominated by spin-coating-based methods.
  \textbf{b} Comparison of selected scalable or industrially relevant fabrication techniques, including slot-die coating, doctor blading, and co-evaporation.
  The plots highlight the strong concentration of high efficiencies at small device areas and the broader area distribution, but increased performance spread, observed for scalable processing routes.}
  \label{fig:scalable_processing}
\end{figure}

The NOMAD interface also enables joint analysis of device performance, active area, and fabrication technique. \Cref{fig:scalable_processing} shows power conversion efficiency as a function of device area, with data points grouped by different deposition and coating methods.

Spin-coating-based processes dominate the literature and account for most high-efficiency devices, which are overwhelmingly reported at small active areas below \SI{1}{\centi\meter\squared}. In contrast, processing routes more closely associated with scalable fabrication, such as slot-die coating and doctor blading, exhibit a broader distribution of device areas, extending to significantly larger devices. However, these larger-area devices show increased performance scatter and generally lower efficiencies.

This example illustrates how NOMAD can be used to contextualize reported efficiencies with respect to fabrication scalability, highlighting the persistent gap between small-area laboratory demonstrations and larger-area devices produced using scalable processing techniques, even if we observe an increase on reports with these techniques.



\printbibliography

%% file: detailed_tables/score_tables.tex
\clearpage

\begin{scriptsize}
\begin{xltabular}{\textwidth}{p{7em} XXXXXXXX}
\caption{Micro-averaged precision aggregated over papers per model.} \\
\toprule
field & scoring method & GPT-4.1 & Claude Opus 4 & GPT-4o & GPT-5 & Claude Sonnet 4 & Claude Opus 4.1 & GPT-5 Mini \\
\midrule
\endfirsthead
\caption[]{Micro-averaged precision aggregated over papers per model} \\
\toprule
field & scoring method & GPT-4.1 & Claude Opus 4 & GPT-4o & GPT-5 & Claude Sonnet 4 & Claude Opus 4.1 & GPT-5 Mini \\
\midrule
\endhead
\midrule
\multicolumn{9}{r}{Continued on next page} \\
\midrule
\endfoot
\bottomrule
\endlastfoot
active\_area& &  &  &  &  &  &  &  \\
\hspace{0.5em}unit & exact match & 0.971 & 0.951 & 0.909 & 0.983 & 0.981 & 0.979 & 0.983 \\
\hspace{0.5em}value &reltol(0.01) & 0.857 & 0.902 & 0.818 & 0.983 & 0.962 & 0.894 & 0.914 \\
averaged\_quantities & type check & 0.852 & 0.912 & 0.679 & 0.889 & 1.000 & 0.914 & 0.526 \\
device\_architecture & exact match & 0.971 & 0.891 & 0.864 & 0.975 & 0.951 & 0.825 & 0.951 \\
encapsulated & type check & 0.769 & 0.769 & 0.750 & 0.942 & 0.949 & 0.864 & 0.556 \\
ff & exact match & 1.000 & 1.000 & 1.000 & 1.000 & 1.000 & 1.000 & 1.000 \\
\hspace{0.5em}value & float | abstol(0.1) & 0.911 & 0.961 & 0.884 & 0.897 & 0.930 & 0.935 & 0.721 \\
jsc& &  &  &  &  &  &  &  \\
\hspace{0.5em}unit & exact match & 1.000 & 1.000 & 0.977 & 1.000 & 1.000 & 1.000 & 0.972 \\
\hspace{0.5em}value & float | abstol(0.1) & 0.946 & 0.961 & 0.953 & 0.985 & 0.958 & 0.952 & 0.847 \\
layers& &  &  &  &  &  &  &  \\
\hspace{0.5em}additional\_treatment & exact match & 0.959 & 0.970 & 0.997 & 0.920 & 0.998 & 0.951 & 0.854 \\
\hspace{0.5em}deposition & exact match & 1.000 & 1.000 & 1.000 & 1.000 & 1.000 & 1.000 & 1.000 \\
\hspace{1.0em}antisolvent & exact match & 1.000 & 0.988 & 1.000 & 0.994 & 0.993 & 0.997 & 1.000 \\
\hspace{1.0em}atmosphere & exact match & 0.824 & 0.865 & 0.753 & 0.465 & 0.943 & 0.914 & 0.644 \\
\hspace{1.0em} duration & &  &  &  &  &  &  &  \\
\hspace{1.5em}unit & exact match & 0.879 & 0.751 & 0.681 & 0.751 & 0.885 & 0.818 & 0.770 \\
\hspace{1.5em}value &reltol(0.01) & 0.648 & 0.716 & 0.507 & 0.652 & 0.793 & 0.737 & 0.645 \\
\hspace{1.0em}method & exact match & 0.895 & 0.904 & 0.816 & 0.890 & 0.919 & 0.869 & 0.753 \\
\hspace{1.0em}solution & exact match & 1.000 & 1.000 & 1.000 & 1.000 & 1.000 & 1.000 & 1.000 \\
\hspace{1.5em}compounds & exact match & NaN & 1.000 & NaN & 1.000 & 1.000 & 1.000 & 1.000 \\
\hspace{1.5em}compounds & nan & NaN & NaN & NaN & NaN & NaN & NaN & NaN \\
\hspace{1.5em}solutes & exact match & NaN & 1.000 & NaN & NaN & 1.000 & 1.000 & 1.000 \\
\hspace{1.0em} solution:solutes & &  &  &  &  &  &  &  \\
\hspace{1.5em} concentration & &  &  &  &  &  &  &  \\
\hspace{2.0em}unit & exact match & NaN & 0.863 & NaN & 0.968 & 0.870 & 1.000 & 1.000 \\
\hspace{2.0em}value &reltol(0.01) & NaN & 0.812 & NaN & 0.903 & 0.840 & 1.000 & 0.931 \\
\hspace{1.5em}name & exact match & NaN & 0.742 & NaN & 0.962 & 0.898 & 0.973 & 0.862 \\
\hspace{1.0em} solution & &  &  &  &  &  &  &  \\
\hspace{1.5em}solvents & exact match & NaN & 1.000 & NaN & NaN & 1.000 & NaN & NaN \\
\hspace{1.0em} solution:solvents & &  &  &  &  &  &  &  \\
\hspace{1.5em}name & exact match & NaN & 0.877 & NaN & 1.000 & 0.976 & 0.900 & 1.000 \\
\hspace{1.5em}volume\_fraction & exact match & NaN & 0.893 & NaN & 0.880 & 0.878 & 0.890 & 1.000 \\
\hspace{1.0em} solution & &  &  &  &  &  &  &  \\
\hspace{1.5em} temperature & &  &  &  &  &  &  &  \\
\hspace{2.0em}unit & type check & NaN & 0.714 & NaN & 0.857 & 0.667 & NaN & 0.000 \\
\hspace{2.0em}value & type check & NaN & 0.714 & NaN & 0.857 & 0.667 & NaN & 0.000 \\
\hspace{1.5em} volume & &  &  &  &  &  &  &  \\
\hspace{2.0em}unit & type check & NaN & 0.000 & NaN & 0.000 & NaN & 0.000 & 0.000 \\
\hspace{2.0em}value & type check & NaN & 0.000 & NaN & 0.000 & NaN & 0.000 & 0.000 \\
\hspace{1.0em}step\_name & type check & 0.668 & 0.653 & 0.797 & 0.487 & 0.623 & 0.639 & 0.438 \\
\hspace{1.0em} temperature & &  &  &  &  &  &  &  \\
\hspace{1.5em}unit & exact match & 0.918 & 0.928 & 0.643 & 1.000 & 0.879 & 0.891 & 0.836 \\
\hspace{1.5em}value &reltol(0.01) & 0.791 & 0.877 & 0.514 & 0.995 & 0.828 & 0.807 & 0.779 \\
\hspace{0.5em}functionality & exact match & 0.904 & 0.927 & 0.920 & 0.943 & 0.985 & 0.898 & 0.917 \\
\hspace{0.5em}name & exact match & 0.919 & 0.961 & 0.917 & 0.927 & 0.987 & 0.946 & 0.865 \\
\hspace{0.5em} thickness & &  &  &  &  &  &  &  \\
\hspace{1.0em}unit & exact match & 0.882 & 0.847 & 1.000 & 0.957 & 0.956 & 0.893 & 0.638 \\
\hspace{1.0em}value &reltol(0.01) & 0.882 & 0.798 & 1.000 & 0.870 & 0.903 & 0.835 & 0.569 \\
light\_source & exact match & 1.000 & 1.000 & 1.000 & NaN & 1.000 & 1.000 & 1.000 \\
\hspace{0.5em}description & type check & 0.041 & 0.848 & 0.000 & 0.028 & 1.000 & 0.821 & 0.029 \\
\hspace{0.5em}lamp & type check & 0.596 & 0.800 & 0.605 & 0.618 & 1.000 & 0.824 & 0.677 \\
\hspace{0.5em} light\_intensity & &  &  &  &  &  &  &  \\
\hspace{1.0em}unit & exact match & 1.000 & 0.951 & 1.000 & 1.000 & 1.000 & 1.000 & 1.000 \\
\hspace{1.0em}value &reltol(0.01) & 1.000 & 0.951 & 1.000 & 1.000 & 1.000 & 1.000 & 1.000 \\
\hspace{0.5em}type & exact match & 1.000 & 1.000 & 0.923 & 0.887 & 0.988 & 1.000 & 0.957 \\
number\_devices &reltol(0.01) & 0.879 & 1.000 & 0.717 & 1.000 & 1.000 & 0.971 & 0.842 \\
pce& &  &  &  &  &  &  &  \\
\hspace{0.5em}unit & exact match & 1.000 & 0.992 & 1.000 & 0.988 & 0.992 & 0.987 & 0.988 \\
\hspace{0.5em}value & float | abstol(0.1) & 0.903 & 0.968 & 0.930 & 0.952 & 0.950 & 0.921 & 0.821 \\
perovskite\_composition& &  &  &  &  &  &  &  \\
\hspace{0.5em}a\_ions & nan & NaN & NaN & NaN & NaN & NaN & NaN & NaN \\
perovskite\_composition:a\_ions& &  &  &  &  &  &  &  \\
\hspace{0.5em}abbreviation & exact match & 0.981 & 0.977 & 0.955 & 0.898 & 1.000 & 0.974 & 0.968 \\
\hspace{0.5em}coefficient & exact match & 0.963 & 0.963 & 0.933 & 0.898 & 1.000 & 0.957 & 0.912 \\
\hspace{0.5em}common\_name & exact match & 0.972 & 0.972 & 1.000 & 0.955 & 0.995 & 0.965 & 0.959 \\
\hspace{0.5em}molecular\_formula & type check & 0.729 & 0.645 & 0.857 & 0.670 & 0.930 & 0.728 & 0.802 \\
perovskite\_composition& &  &  &  &  &  &  &  \\
\hspace{0.5em}b\_ions & nan & NaN & NaN & NaN & NaN & NaN & NaN & NaN \\
perovskite\_composition:b\_ions& &  &  &  &  &  &  &  \\
\hspace{0.5em}abbreviation & exact match & 1.000 & 1.000 & 1.000 & 1.000 & 1.000 & 1.000 & 1.000 \\
\hspace{0.5em}coefficient & exact match & 1.000 & 1.000 & 1.000 & 1.000 & 1.000 & 1.000 & 1.000 \\
\hspace{0.5em}common\_name & exact match & 0.986 & 0.992 & 1.000 & 0.986 & 0.992 & 0.986 & 0.988 \\
\hspace{0.5em}molecular\_formula & type check & 0.971 & 0.992 & 0.960 & 0.985 & 0.992 & 0.986 & 0.986 \\
perovskite\_composition& &  &  &  &  &  &  &  \\
\hspace{0.5em} bandgap & &  &  &  &  &  &  &  \\
\hspace{1.0em}unit & exact match & 0.667 & 0.775 & 0.788 & 0.765 & 0.827 & 0.877 & 0.829 \\
\hspace{1.0em}value &reltol(0.01) & 0.417 & 0.648 & 0.667 & 0.662 & 0.773 & 0.785 & 0.610 \\
\hspace{0.5em}dimensionality & type check & 0.847 & 0.922 & 0.881 & 0.892 & 0.927 & 0.877 & 0.847 \\
\hspace{0.5em}formula & exact match & 0.972 & 0.969 & 0.898 & 0.964 & 0.992 & 0.950 & 0.847 \\
\hspace{0.5em}x\_ions & nan & NaN & NaN & NaN & NaN & NaN & NaN & NaN \\
perovskite\_composition:x\_ions& &  &  &  &  &  &  &  \\
\hspace{0.5em}abbreviation & exact match & 1.000 & 1.000 & 1.000 & 1.000 & 1.000 & 1.000 & 1.000 \\
\hspace{0.5em}coefficient & exact match & 0.949 & 0.935 & 0.775 & 0.893 & 0.954 & 0.912 & 0.835 \\
\hspace{0.5em}common\_name & exact match & 0.990 & 0.995 & 1.000 & 0.991 & 0.994 & 0.991 & 0.991 \\
\hspace{0.5em}molecular\_formula & type check & 0.980 & 0.995 & 0.964 & 0.990 & 0.994 & 0.991 & 0.990 \\
stability & exact match & 1.000 & 1.000 & 1.000 & 1.000 & 1.000 & 1.000 & 1.000 \\
\hspace{0.5em} humidity & &  &  &  &  &  &  &  \\
\hspace{1.0em}unit & exact match & 0.667 & 0.400 & 1.000 & NaN & 0.800 & 1.000 & 1.000 \\
\hspace{1.0em}value &reltol(0.01) & 0.667 & 0.400 & 1.000 & NaN & 0.800 & 1.000 & 1.000 \\
\hspace{0.5em} light\_intensity & &  &  &  &  &  &  &  \\
\hspace{1.0em}unit & exact match & 1.000 & 1.000 & NaN & 1.000 & 1.000 & 1.000 & 0.667 \\
\hspace{1.0em}value &reltol(0.01) & 1.000 & 1.000 & NaN & 1.000 & 1.000 & 1.000 & 0.667 \\
\hspace{0.5em} PCE\_after\_\_hours & &  &  &  &  &  &  &  \\
\hspace{1.0em}unit & exact match & 1.000 & 1.000 & 0.625 & NaN & 1.000 & 1.000 & 1.000 \\
\hspace{1.0em}value &reltol(0.01) & 0.667 & 0.800 & 0.625 & NaN & 0.750 & 1.000 & NaN \\
\hspace{0.5em} PCE\_at\_the\_end\_of\_description & &  &  &  &  &  &  &  \\
\hspace{1.0em}unit & nan & NaN & NaN & NaN & NaN & NaN & NaN & NaN \\
\hspace{1.0em}value & nan & NaN & NaN & NaN & NaN & NaN & NaN & NaN \\
\hspace{0.5em} PCE\_at\_the\_start\_of\_the\_experiment & &  &  &  &  &  &  &  \\
\hspace{1.0em}unit & exact match & 1.000 & 0.857 & 1.000 & NaN & 1.000 & 1.000 & 1.000 \\
\hspace{1.0em}value &reltol(0.01) & 1.000 & 0.786 & 0.909 & NaN & 1.000 & 1.000 & 0.750 \\
\hspace{0.5em} PCE\_T & &  &  &  &  &  &  &  \\
\hspace{1.0em}unit & exact match & 0.500 & 1.000 & NaN & 0.000 & 1.000 & 1.000 & 1.000 \\
\hspace{1.0em}value &reltol(0.01) & 0.500 & 1.000 & NaN & 0.000 & 1.000 & 1.000 & 1.000 \\
\hspace{0.5em}potential\_bias & exact match & 1.000 & 0.750 & 0.727 & 1.000 & 1.000 & 1.000 & 0.750 \\
\hspace{0.5em} temperature & &  &  &  &  &  &  &  \\
\hspace{1.0em}unit & exact match & 1.000 & 0.625 & 1.000 & NaN & 0.875 & 1.000 & 1.000 \\
\hspace{1.0em}value &reltol(0.01) & 1.000 & 0.500 & 1.000 & NaN & 0.875 & 1.000 & 1.000 \\
\hspace{0.5em} time & &  &  &  &  &  &  &  \\
\hspace{1.0em}unit & exact match & 1.000 & 1.000 & NaN & 1.000 & 1.000 & 1.000 & 1.000 \\
\hspace{1.0em}value &reltol(0.01) & 0.800 & 0.786 & NaN & 1.000 & 1.000 & 0.875 & 0.667 \\
voc& &  &  &  &  &  &  &  \\
\hspace{0.5em}unit & exact match & 1.000 & 0.987 & 0.930 & 1.000 & 0.986 & 0.985 & 0.973 \\
\hspace{0.5em}value & float | abstol(0.01) & 0.946 & 0.949 & 0.930 & 0.958 & 0.944 & 0.938 & 0.849 \\
\end{xltabular}
\end{scriptsize}

\begin{scriptsize}
\begin{xltabular}{\textwidth}{p{7em} XXXXXXXX}
\caption{Micro-averaged recall aggregated over papers per model.} \\
\toprule
field & scoring method & GPT-4.1 & Claude Opus 4 & GPT-4o & GPT-5 & Claude Sonnet 4 & Claude Opus 4.1 & GPT-5 Mini \\
\midrule
\endfirsthead
\caption[]{Micro-averaged recall aggregated over papers per model.} \\
\toprule
field & scoring method & GPT-4.1 & Claude Opus 4 & GPT-4o & GPT-5 & Claude Sonnet 4 & Claude Opus 4.1 & GPT-5 Mini \\
\midrule
\endhead
\midrule
\multicolumn{9}{r}{Continued on next page} \\
\midrule
\endfoot
\bottomrule
\endlastfoot
active\_area & &  &  &  &  &  &  &  \\
\hspace{0.5em}unit & exact match & 0.252 & 0.436 & 0.149 & 0.437 & 0.385 & 0.341 & 0.422 \\
\hspace{0.5em}value &reltol(0.01) & 0.229 & 0.423 & 0.136 & 0.437 & 0.381 & 0.321 & 0.405 \\
averaged\_quantities & type check & 0.409 & 0.817 & 0.303 & 0.500 & 0.846 & 0.492 & 0.414 \\
device\_architecture & exact match & 0.500 & 0.934 & 0.398 & 0.590 & 0.900 & 0.541 & 0.583 \\
encapsulated & type check & 0.413 & 0.748 & 0.317 & 0.492 & 0.862 & 0.398 & 0.450 \\
ff & exact match & 0.239 & 0.978 & 0.304 & 0.065 & 0.978 & 0.174 & 0.087 \\
\hspace{0.5em}value & float | abstol(0.1) & 0.600 & 0.839 & 0.447 & 0.735 & 0.776 & 0.674 & 0.690 \\
jsc & &  &  &  &  &  &  &  \\
\hspace{0.5em}unit & exact match & 0.412 & 0.559 & 0.311 & 0.500 & 0.522 & 0.463 & 0.522 \\
\hspace{0.5em}value & float | abstol(0.1) & 0.398 & 0.549 & 0.306 & 0.496 & 0.511 & 0.451 & 0.488 \\
layers & &  &  &  &  &  &  &  \\
\hspace{0.5em}additional\_treatment & exact match & 0.490 & 0.891 & 0.397 & 0.570 & 0.848 & 0.536 & 0.492 \\
\hspace{0.5em}deposition & exact match & 1.000 & 1.000 & 1.000 & 1.000 & 1.000 & 1.000 & 1.000 \\
\hspace{1.0em}antisolvent & exact match & 0.503 & 0.523 & 0.339 & 0.748 & 0.668 & 0.470 & 0.596 \\
\hspace{1.0em}atmosphere & exact match & 0.354 & 0.486 & 0.306 & 0.633 & 0.656 & 0.387 & 0.555 \\
\hspace{1.0em} duration & &  &  &  &  &  &  &  \\
\hspace{1.5em}unit & exact match & 0.130 & 0.217 & 0.078 & 0.398 & 0.254 & 0.186 & 0.198 \\
\hspace{1.5em}value &reltol(0.01) & 0.099 & 0.209 & 0.059 & 0.365 & 0.234 & 0.171 & 0.171 \\
\hspace{1.0em}method & exact match & 0.472 & 0.523 & 0.318 & 0.767 & 0.650 & 0.441 & 0.594 \\
\hspace{1.0em}solution & exact match & 0.614 & 0.734 & 0.373 & 0.760 & 0.854 & 0.562 & 0.695 \\
\hspace{1.5em}compounds & exact match & 0.000 & 0.072 & 0.000 & 0.177 & 0.351 & 0.095 & 0.033 \\
\hspace{1.5em}compounds & nan & 0.000 & 0.000 & 0.000 & 0.000 & 0.000 & 0.000 & 0.000 \\
\hspace{1.5em}solutes & exact match & 0.000 & 0.167 & 0.000 & 0.000 & 0.292 & 0.125 & 0.083 \\
\hspace{1.0em} solution:solutes & &  &  &  &  &  &  &  \\
\hspace{1.5em} concentration & &  &  &  &  &  &  &  \\
\hspace{2.0em}unit & exact match & 0.000 & 0.140 & 0.000 & 0.180 & 0.365 & 0.147 & 0.058 \\
\hspace{2.0em}value &reltol(0.01) & 0.000 & 0.133 & 0.000 & 0.170 & 0.357 & 0.147 & 0.054 \\
\hspace{1.5em}name & exact match & 0.000 & 0.138 & 0.000 & 0.202 & 0.384 & 0.144 & 0.050 \\
\hspace{1.0em} solution & &  &  &  &  &  &  &  \\
\hspace{1.5em}solvents & exact match & 0.000 & 0.125 & 0.000 & 0.000 & 0.375 & 0.000 & 0.000 \\
\hspace{1.0em} solution:solvents & &  &  &  &  &  &  &  \\
\hspace{1.5em}name & exact match & 0.000 & 0.246 & 0.000 & 0.236 & 0.399 & 0.197 & 0.102 \\
\hspace{1.5em}volume\_fraction & exact match & 0.000 & 0.225 & 0.000 & 0.106 & 0.323 & 0.158 & 0.024 \\
\hspace{1.0em} solution & &  &  &  &  &  &  &  \\
\hspace{1.5em} temperature & &  &  &  &  &  &  &  \\
\hspace{2.0em}unit & type check & 0.000 & 0.013 & 0.000 & 0.015 & 0.021 & 0.000 & 0.000 \\
\hspace{2.0em}value & type check & 0.000 & 0.013 & 0.000 & 0.015 & 0.021 & 0.000 & 0.000 \\
\hspace{1.5em} volume & &  &  &  &  &  &  &  \\
\hspace{2.0em}unit & type check & 0.000 & 0.000 & 0.000 & 0.000 & 0.000 & 0.000 & 0.000 \\
\hspace{2.0em}value & type check & 0.000 & 0.000 & 0.000 & 0.000 & 0.000 & 0.000 & 0.000 \\
\hspace{1.0em}step\_name & type check & 0.442 & 0.371 & 0.205 & 0.643 & 0.558 & 0.332 & 0.399 \\
\hspace{1.0em} temperature & &  &  &  &  &  &  &  \\
\hspace{1.5em}unit & exact match & 0.164 & 0.208 & 0.075 & 0.297 & 0.253 & 0.173 & 0.168 \\
\hspace{1.5em}value &reltol(0.01) & 0.144 & 0.199 & 0.061 & 0.296 & 0.242 & 0.159 & 0.159 \\
\hspace{0.5em}functionality & exact match & 0.491 & 0.899 & 0.396 & 0.594 & 0.866 & 0.536 & 0.518 \\
\hspace{0.5em}name & exact match & 0.495 & 0.902 & 0.395 & 0.590 & 0.866 & 0.549 & 0.503 \\
\hspace{0.5em} thickness & &  &  &  &  &  &  &  \\
\hspace{1.0em}unit & exact match & 0.019 & 0.137 & 0.023 & 0.085 & 0.139 & 0.119 & 0.048 \\
\hspace{1.0em}value &reltol(0.01) & 0.019 & 0.130 & 0.023 & 0.077 & 0.132 & 0.112 & 0.043 \\
light\_source & exact match & 0.171 & 1.000 & 0.244 & 0.000 & 1.000 & 0.098 & 0.024 \\
\hspace{0.5em}description & type check & 0.042 & 0.807 & 0.000 & 0.077 & 0.821 & 0.663 & 0.074 \\
\hspace{0.5em}lamp & type check & 0.368 & 0.578 & 0.287 & 0.609 & 0.621 & 0.488 & 0.595 \\
\hspace{0.5em} light\_intensity & &  &  &  &  &  &  &  \\
\hspace{1.0em}unit & exact match & 0.242 & 0.846 & 0.126 & 0.611 & 0.811 & 0.674 & 0.600 \\
\hspace{1.0em}value &reltol(0.01) & 0.242 & 0.846 & 0.126 & 0.611 & 0.811 & 0.674 & 0.600 \\
\hspace{0.5em}type & exact match & 0.516 & 0.853 & 0.391 & 0.724 & 0.840 & 0.726 & 0.728 \\
number\_devices &reltol(0.01) & 0.453 & 0.853 & 0.314 & 0.493 & 0.809 & 0.507 & 0.516 \\
pce & &  &  &  &  &  &  &  \\
\hspace{0.5em}unit & exact match & 0.529 & 0.919 & 0.419 & 0.607 & 0.889 & 0.556 & 0.615 \\
\hspace{0.5em}value & float | abstol(0.1) & 0.504 & 0.917 & 0.402 & 0.598 & 0.885 & 0.538 & 0.570 \\
perovskite\_composition & &  &  &  &  &  &  &  \\
\hspace{0.5em}a\_ions & nan & 0.000 & 0.000 & 0.000 & 0.000 & 0.000 & 0.000 & 0.000 \\
perovskite\_composition:a\_ions & &  &  &  &  &  &  &  \\
\hspace{0.5em}abbreviation & exact match & 0.465 & 0.937 & 0.379 & 0.535 & 0.934 & 0.498 & 0.540 \\
\hspace{0.5em}coefficient & exact match & 0.460 & 0.936 & 0.374 & 0.535 & 0.934 & 0.493 & 0.525 \\
\hspace{0.5em}common\_name & exact match & 0.462 & 0.937 & 0.123 & 0.475 & 0.934 & 0.496 & 0.520 \\
\hspace{0.5em}molecular\_formula & type check & 0.392 & 0.908 & 0.107 & 0.356 & 0.930 & 0.421 & 0.348 \\
perovskite\_composition & &  &  &  &  &  &  &  \\
\hspace{0.5em}b\_ions & nan & 0.000 & 0.000 & 0.000 & 0.000 & 0.000 & 0.000 & 0.000 \\
perovskite\_composition:b\_ions & &  &  &  &  &  &  &  \\
\hspace{0.5em}abbreviation & exact match & 0.522 & 0.940 & 0.425 & 0.604 & 0.903 & 0.575 & 0.612 \\
\hspace{0.5em}coefficient & exact match & 0.522 & 0.940 & 0.425 & 0.604 & 0.903 & 0.575 & 0.612 \\
\hspace{0.5em}common\_name & exact match & 0.519 & 0.917 & 0.187 & 0.511 & 0.902 & 0.549 & 0.594 \\
\hspace{0.5em}molecular\_formula & type check & 0.515 & 0.917 & 0.180 & 0.489 & 0.902 & 0.549 & 0.534 \\
perovskite\_composition & &  &  &  &  &  &  &  \\
\hspace{0.5em} bandgap & &  &  &  &  &  &  &  \\
\hspace{1.0em}unit & exact match & 0.429 & 0.458 & 0.202 & 0.433 & 0.504 & 0.445 & 0.264 \\
\hspace{1.0em}value &reltol(0.01) & 0.319 & 0.414 & 0.176 & 0.398 & 0.487 & 0.418 & 0.208 \\
\hspace{0.5em}dimensionality & type check & 0.488 & 0.944 & 0.403 & 0.583 & 0.898 & 0.563 & 0.585 \\
\hspace{0.5em}formula & exact match & 0.522 & 0.939 & 0.408 & 0.602 & 0.904 & 0.576 & 0.585 \\
\hspace{0.5em}x\_ions & nan & 0.000 & 0.000 & 0.000 & 0.000 & 0.000 & 0.000 & 0.000 \\
perovskite\_composition:x\_ions & &  &  &  &  &  &  &  \\
\hspace{0.5em}abbreviation & exact match & 0.493 & 0.920 & 0.398 & 0.607 & 0.866 & 0.562 & 0.602 \\
\hspace{0.5em}coefficient & exact match & 0.480 & 0.915 & 0.339 & 0.580 & 0.860 & 0.539 & 0.558 \\
\hspace{0.5em}common\_name & exact match & 0.490 & 0.905 & 0.139 & 0.535 & 0.865 & 0.545 & 0.580 \\
\hspace{0.5em}molecular\_formula & type check & 0.487 & 0.905 & 0.135 & 0.515 & 0.865 & 0.545 & 0.475 \\
stability & exact match & 0.346 & 0.506 & 0.272 & 0.358 & 0.951 & 0.333 & 0.346 \\
\hspace{0.5em} humidity & &  &  &  &  &  &  &  \\
\hspace{1.0em}unit & exact match & 0.037 & 0.038 & 0.018 & 0.000 & 0.074 & 0.036 & 0.018 \\
\hspace{1.0em}value &reltol(0.01) & 0.037 & 0.038 & 0.018 & 0.000 & 0.074 & 0.036 & 0.018 \\
\hspace{0.5em} light\_intensity & &  &  &  &  &  &  &  \\
\hspace{1.0em}unit & exact match & 0.055 & 0.127 & 0.000 & 0.036 & 0.109 & 0.109 & 0.037 \\
\hspace{1.0em}value &reltol(0.01) & 0.055 & 0.127 & 0.000 & 0.036 & 0.109 & 0.109 & 0.037 \\
\hspace{0.5em} PCE\_after\_\_hours & &  &  &  &  &  &  &  \\
\hspace{1.0em}unit & exact match & 0.055 & 0.091 & 0.096 & 0.000 & 0.145 & 0.073 & 0.018 \\
\hspace{1.0em}value &reltol(0.01) & 0.037 & 0.074 & 0.096 & 0.000 & 0.113 & 0.073 & 0.000 \\
\hspace{0.5em} PCE\_at\_the\_end\_of\_description & &  &  &  &  &  &  &  \\
\hspace{1.0em}unit & nan & 0.000 & 0.000 & 0.000 & 0.000 & 0.000 & 0.000 & 0.000 \\
\hspace{1.0em}value & nan & 0.000 & 0.000 & 0.000 & 0.000 & 0.000 & 0.000 & 0.000 \\
\hspace{0.5em} PCE\_at\_the\_start\_of\_the\_experiment & &  &  &  &  &  &  &  \\
\hspace{1.0em}unit & exact match & 0.091 & 0.226 & 0.200 & 0.000 & 0.236 & 0.145 & 0.073 \\
\hspace{1.0em}value &reltol(0.01) & 0.091 & 0.212 & 0.185 & 0.000 & 0.236 & 0.145 & 0.056 \\
\hspace{0.5em} PCE\_T & &  &  &  &  &  &  &  \\
\hspace{1.0em}unit & exact match & 0.019 & 0.036 & 0.000 & 0.000 & 0.036 & 0.018 & 0.018 \\
\hspace{1.0em}value &reltol(0.01) & 0.019 & 0.036 & 0.000 & 0.000 & 0.036 & 0.018 & 0.018 \\
\hspace{0.5em}potential\_bias & exact match & 0.091 & 0.113 & 0.154 & 0.036 & 0.236 & 0.109 & 0.056 \\
\hspace{0.5em} temperature & &  &  &  &  &  &  &  \\
\hspace{1.0em}unit & exact match & 0.018 & 0.096 & 0.018 & 0.000 & 0.130 & 0.055 & 0.018 \\
\hspace{1.0em}value &reltol(0.01) & 0.018 & 0.078 & 0.018 & 0.000 & 0.130 & 0.055 & 0.018 \\
\hspace{0.5em} time & &  &  &  &  &  &  &  \\
\hspace{1.0em}unit & exact match & 0.091 & 0.255 & 0.000 & 0.018 & 0.236 & 0.145 & 0.055 \\
\hspace{1.0em}value &reltol(0.01) & 0.074 & 0.212 & 0.000 & 0.018 & 0.236 & 0.130 & 0.037 \\
voc & &  &  &  &  &  &  &  \\
\hspace{0.5em}unit & exact match & 0.412 & 0.570 & 0.301 & 0.522 & 0.519 & 0.474 & 0.530 \\
\hspace{0.5em}value & float | abstol(0.01) & 0.398 & 0.561 & 0.301 & 0.511 & 0.508 & 0.462 & 0.496 \\
\end{xltabular}
\end{scriptsize}